\documentclass[prb,twocolumn,prb,showpacs,aps]{revtex4}
\usepackage{amsmath,amsfonts,bm,graphicx,dsfont}
\usepackage{epsfig}
\usepackage{amssymb}

\newcommand{\be}{\begin{equation}}
\newcommand{\ee}{\end{equation}}

\renewcommand{\d}{\text{d}}

\usepackage{color}

\begin{document}

\title{Transport across an Anderson quantum dot in the intermediate coupling regime}
\author{Johannes Kern}
\email{johannes.kern@physik.uni-regensburg.de}
\author{Milena Grifoni}
\author{}
 
\affiliation{Institut f\"ur Theoretische Physik, Universit\"at Regensburg, 93040 Regensburg, Germany}
\date{\today}

\begin{abstract}
We describe linear and nonlinear transport across a single impurity Anderson model quantum dot with 
intermediate coupling to the leads, i.e., with tunnel coupling of the order of the thermal energy 
$k_B T$. The coupling is large enough that
sequential tunneling processes alone do not suffice to properly describe the transport characteristics. Upon 
applying  a density matrix approach, the current is expressed in terms of rates obtained by considering a very
small class of diagrams which dress the sequential tunneling processes by charge fluctuations.
We call this the ``dressed second order'' (DSO) approximation. One major achievement of the DSO is that, still
in the Coulomb blockade regime, it can describe the crossover from thermally broadened to tunneling broadened 
conductance peaks. 
When the temperature is decreased even further, the DSO captures ``Kondesque'' behaviours of the Anderson 
quantum dot qualitatively: 
We find a zero bias anomaly of the differential conductance versus applied bias, an
enhancement of the conductance with decreasing temperature as well as the onset of universality  of the shape 
of the conductance as function of the temperature. We can address the case of a spin-degenerate level split
energetically by a magnetic field and show that, if we assume in addition different capacitive couplings of the 
two spin-levels to the leads, one of the resonance peaks is vanishing. In case spin-dependent
chemical potentials are introduced and only one of the four is varied, the DSO yields {\em in principle} only one 
resonance. This seems to be in agreement with experiments with pseudo-spin \cite{Wilhelm02}.    
Furthermore, we get qualitative agreement with experimental data showing a cross-over from the Kondo to 
the empty orbital regime.
\end{abstract}


\maketitle

\section{Introduction}
The single impurity Anderson model (SIAM) \cite{Anderson61} has become a useful tool to describe  
phenomena arising in quantum dot
devices at low temperatures. It encompasses single-electron tunneling phenomena \cite{Grabert92}, cotunneling  
and resonant tunneling \cite{Sohn96}
as well as Kondo \cite{Kondo64} physics.
These phenomena have been verified in many experimental quantum dot set-ups 
realized at the interface of a two-dimensional electron gas 
\cite{Goldhaber98Nature,Goldhaber98,Schmid98,Cronenwett98,Wiel00,Grobis08},
 carbon nanotubes \cite{Nygard00,Herrero05,Sahoo05,Hauptmann08,Gaass11} and quantum wires 
\cite{Csonka08,Kretinin11} as well as in single-molecule junctions \cite{Pasupathy04}.
At thermal energies larger than the coupling $\Gamma$ to the leads transport can be 
suppressed 
at low bias due to Coulomb blockade. In this regime  sequential tunneling dominates the transport
across  the SIAM, and tunneling in and out of the dot is well described in terms of rate equations 
\cite{Averin91,Beenakker91}, with rates obtained within a second order perturbation theory in the tunneling 
Hamiltonian (i.e., first order in $\Gamma$). When the temperature is decreased to values of the order of the 
tunneling coupling $\Gamma$ or lower, the sequential tunneling approximation breaks down, as processes of higher
order in $\Gamma$ start to become important.

The intermediate coupling regime with tunnel couplings of the order of the thermal energy or higher has not been 
much investigated so far. This is in part due to the difficulty of developing  theories capable to cope 
with strong Coulomb interactions and intermediate coupling at the same time.
However, it is this intermediate regime which might be of interest for transport through some 
single-molecule junctions \cite{Reed, Lörtscher} and is relevant to interpret experiments on negative tunneling 
magnetoresistance \cite{Sahoo05}. In the single molecule experiments \cite{Reed, Lörtscher} a conductance gap is 
observed at low bias which suggests the presence of charging effects. The gap is followed by conductance peaks 
whose broadening is larger than the estimated temperature, being a hint that tunneling processes of high order 
might be responsible for the broadening. In Ref. \cite{Sahoo05} Coulomb oscillations of the conductance versus the 
gate voltage are clearly seen in carbon nanotubes contacted to ferromagnetic leads; however, the occurrence of  a 
negative magnetoresistance requires the presence of level shifts due to higher order charge fluctuation processes 
\cite{Cottet, Koller12}.

When the temperature is increased even further, one observes the  
occurrence of a zero bias maximum \cite{Glazman88,Ng88,Meir93,Wingreen94,Ralph94,Schmid98} or minimum 
\cite{König96,Schmid98} of the nonlinear conductance for  small temperatures in a quantum dot with large 
Coulomb interaction, 
depending on whether the single particle resonance lies deep below (Kondo regime) or above 
(empty-orbital regime) the Fermi level, respectively.

In this paper we describe the transport beyond the sequential tunneling regime by using a diagrammatic approach 
to the stationary reduced density 
matrix of the quantum dot and the stationary electron current onto one of the leads. 
Along the same lines as in Ref.\cite{Koller12} we include all possible diagrams which dress the second order 
tunneling rates by charge fluctuations in and out of the quantum dot. Different from the method in  
\cite{Koller12}, we do not only extract tunneling induced level shifts from the analytical expressions.  
 We calculate {\em transition rates} and express the density matrix and the current in terms of those. 
Our "dressed second order" (DSO) diagrams are a small subset of the diagrams kept within the so called 
resonant tunneling approximation (RTA), first proposed by \cite{Schoeller94} to describe the beyond sequential 
tunneling regime. In particular, for a spinless quantum dot the RTA is exact and reproduces e.g. 
the expected Breit-Wigner resonance shape of the linear conductance. The much smaller 
DSO subset, too,  yields the known exact result for the current.

The DSO yields the rates in a straightforward way.
We compared the predictions of the RTA and DSO both for the linear and nonlinear conductance in the case of 
infinitely large Coulomb-repulsion and found 
only small deviations in the intermediate coupling regime.
Larger deviations are seen at lower temperatures where the conductance obtained by the DSO is remaining 
considerably below the RTA-result.

One major achievement of the DSO is its capability to properly describe a cross-over from thermally broadened
conductance peaks at high temperatures to tunneling broadened conductance peaks at low temperatures. This is 
of relevance e.g. to explain the experiments of Ref. \cite{Sahoo05}. The DSO tunneling rates are given in 
integral form with the integrand including the product of the density of electron levels and a 
Lorentzian-like function. Interestingly, a similar form is necessary to ensure convergence of the current in 
models of quantum dots coupled to superconducting leads \cite{Levy97}. Hence, the DSO also provides the minimum
diagram selection which yields effective Dynes spectral densities \cite{Pekola10} in superconducting set-ups.

For small temperatures one expects a zero bias resonance in the transport across a quantum dot with odd 
occupation and large Coulomb interaction. Both the DSO and the RTA contain this resonance. To test the reliability
of the DSO we thus investigated the temperature dependence of the linear conductance obtained by it. We found 
that there is a temperature 
$T_K$ such  that the conductance is a universal function of the ratio $T/T_K$.
We compared our expression for $T_K$ with results of other theories for the Kondo temperature in the case of 
infinite Coulomb interaction. Interestingly, as in the RTA \cite{Schoeller97}, we find an exponent differing 
precisely by a factor of two from the result in Ref. \cite{Bickers87}. Moreover, the shape of the conductance 
curve differs from that expected e.g. from numerical renormalization group predictions.

To show the predictive power of the DSO on a {\em qualitative } level we
address the case that the two spin levels are split energetically by a magnetic field and reproduce 
the result that the zero bias resonance of the conductance versus the bias splits up into two peaks
\cite{Meir93}.
Moreover, we consider the situation that in addition to an energy difference the levels have different 
capacitive couplings to the contacts (still equal tunneling coupling). In this case we obtain that one part 
of the double peak vanishes with increasing asymmetry. 
Measurements which might be explained by this effect were reported in Ref. \cite{Schmid98}.
Furthermore, we show the behaviour of the DSO-resonance in an unconventional situation: The chemical 
potentials of the leads depend on the spin. Only one of the four chemical potentials is varied and 
the others are kept constant and equal. In this situation, the DSO yields only one resonance. Experimental results
with a pseudo-spin instead of real spin \cite{Wilhelm02} seem to be in agreement with this prediction.

Finally, we focus on the case of finite but still large Coulomb interaction and consider the linear conductance 
as a function of the gate voltage and the temperature. The effect of changing the gate voltage is a shift of the 
relative position
of the level energy with respect to the Fermi level. 
This offers the possibility to investigate the cross-over from the Kondo regime 
to the mixed valence  and finally the empty orbital regime, corresponding to single particle energies 
lying deep below, in the vicinity (within an uncertainty of the order of  $\Gamma$) and above 
the Fermi level of the leads, respectively. We compare with experimental results in Ref. \cite{Goldhaber98}
and obtain in many respects qualitative agreement. 

In summary, despite its simplicity, the DSO provides important insights on the physics of a correlated Anderson 
quantum dot over a broad regime of parameters. Because the theory is easily scalable to multilevel quantum dots 
set-ups, it could become an interesting tool to investigate complex quantum dot systems.

The structure of the paper is as follows: Section II introduces the model of the transport current. 
Section III illustrates the diagrammatic approach and recalls known results for the reduced density 
matrix and the current in second order in the tunneling Hamiltonian. Analytical expressions for the current 
and the reduced density matrix are provided in terms of rates.

The DSO approximation is 
explained in Sec. IV.   In Sec. V, VI, and VII the DSO is applied to the spinless case, to the case of 
infinite interaction and finite interaction,
respectively. In particular, in Sec. V and VI the DSO and RTA predictions are compared; the case of 
energetically split levels is considered. 
In Sec. VII 
on the other hand we compare with the experimental results in \cite{Goldhaber98}. Finally, conclusions 
are drawn in Sec. VIII.

\section{Basic model}

\subsection{Hamiltonian}
The Hamilton operator of our system is $H = H_R + H_\odot + H_T$. In the reservoirs we assume noninteracting 
electrons. Correspondingly, we choose 
\begin{displaymath}
 H_R = \sum_{l\sigma {\bf k}} \varepsilon_{l \sigma {\bf k}}  c_{l \sigma {\bf k}}^\dagger c_{l \sigma {\bf k}}.
\end{displaymath}
In this formula, the indices $l$, $\sigma$ and ${\bf k} $ denote the lead, the spin and the wave vector of an
electron level in the contacts, respectively; $\varepsilon_{l \sigma {\bf k}}$ is the band energy corresponding to 
this electron level;  $c_{l \sigma {\bf k}}$ is the annihilation operator of the level {$l \sigma {\bf k}$} 
and the dagger denotes the Hermitian conjugate.

The Hamiltonian of the isolated quantum dot is 
\begin{displaymath}
 H_\odot = U d_\uparrow^\dagger \d_\uparrow  d_\downarrow^\dagger d_\downarrow  + \sum_\sigma E_\sigma 
  d_\sigma^\dagger d_\sigma,   
\end{displaymath}
where $U$ is the Coulomb interaction and $d_\sigma$  and $d_\sigma^\dagger$ are the annihilation and creation 
operator of the level $\sigma = \uparrow / \downarrow$ on the dot. Alternatively, the Hamiltonian of the isolated
dot can be written as     
\begin{displaymath}
H_\odot = E_0  |0\rangle \langle 0| + \sum_\sigma E_\sigma |\sigma \rangle \langle \sigma| + E_2 |2\rangle  
\langle 2|. 
\end{displaymath}
For any of the four many particle states  $a = 0, \uparrow, \downarrow, 2$ we use ``$E_a$'' to denote the energy 
of this state. The Hamiltonian of the quantum dot is diagonal in the basis given by these four states. By
comparison with the above representation we get: $E_0 = 0, E_2 = U + \sum_\sigma E_\sigma$. Only differences 
between energies of quantum dot states are relevant. We introduce the terminology
\begin{displaymath}
 E_{ab} := E_a - E_b.
\end{displaymath}

Finally, the tunneling Hamiltonian, 
\begin{displaymath} \label{H_T}
 H_T = \sum_{ l \sigma {\bf k}} T_{l{\bf k}\sigma}  d_{\sigma}^\dagger c_{l{\bf k}\sigma} + 
		      \mbox{H. c. (Hermitian conjugate)},
\end{displaymath}
connects electron levels on the leads with the level on the quantum dot \cite{Bardeen}. With the term 
``order'' of a tunneling process/ diagram we mean its order in the 
``tunneling coupling'' or ''in $H_T$``.

\subsection{Initial condition}
We assume that there is an initial time at which the systems are still separate and express this by 
writing the initial density matrix as product of density matrices of the quantum dot and the leads: 
\begin{displaymath}\label{initial density matrix}
 \rho (t_0) = \rho_{\odot} (t_0) \otimes \rho_R,
\end{displaymath}
where $\rho_{\odot} (t_0)$ is some arbitrary initial density matrix describing the state of the dot; 
$\rho_R = \rho_{R,left} \otimes \rho_{R,right} $ is the density matrix of the leads in thermal equilibrium. 
Specifically,  we choose
\begin{displaymath}
\rho_{R,l} =  \frac{1}{n_l} \exp \left( \frac{-1}{k_B T} \sum_{{\bf k}\sigma} (\varepsilon_{l{\bf k}\sigma}
  - \mu_l)  c_{l{\bf k}\sigma}^\dagger c_{l{\bf k}\sigma} \right),
\end{displaymath}
where $\mu_l$ is the chemical potential of lead $l$ and where $n_l$ is a normalization factor. After this 
initial time we assume that the time evolution of the density matrix is ruled by the total Hamiltonian $H$ 
according to the Liouville-von Neumann equation \cite{Blum} which is the analogon of the Schr{\"o}dinger 
equation for density matrices.

\subsection{Thermodynamic limit and the current}
For each of the leads we define an electron counting operator as $N_l = \sum_{{\bf k} \sigma} 
c_{l{\bf k} \sigma}^\dagger c_{l{\bf k} \sigma}$ and the operator of the particle current onto that lead 
as $I_l =  \frac{i}{\hbar} [H, N_l]$. Then the current onto the chosen lead at time $t$ is  
\begin{displaymath}
 \frac{d}{dt} trace \left( N_l \rho(t) \right) = Tr \left( I_l \rho(t) \right) =: < I_l > (t). 
\end{displaymath}
We define the stationary current by letting the time go to infinity and taking the average current: 
\begin{equation} \label{current}
 <I_l>_\infty = \lim_{\lambda \to 0} \lambda \int_{t_0}^\infty dt < I_l > (t) e^{-\lambda (t- t_0)}, 
\end{equation}
where $\lambda > 0$ is the argument of the Laplace transform of the function $< I_l >(t)$. The total 
weight of the multiplicant $\lambda e^{-\lambda (t - t_0)}$ over $ [t_0, \infty[$ is always unity, but for 
smaller and smaller values of $\lambda$ it will be distributed over a larger and larger time interval. 
The current in this definition is zero as long as the contacts are finite. Therefore, we  
first let the size of the contacts go to infinity and redefine
\begin{equation} \label{current(t)}
 <I_l> (t) = \lim_{V \to \infty} <I_l> (t, V). 
\end{equation}
Then the current in the definition of Eq. (\ref{current}) is our model of the dc-current measured in 
transport experiments with quantum dots.

\section{Diagrammatic approach}
\subsection{Basic method}
An analysis of the time evolution of the current, Eq. (\ref{current(t)}), shows that it can be separated into 
subsequent smallest segments, so-called irreducible tunneling processes \cite{Schoeller97, Report10}. The 
calculation of the 
stationary current can be reduced to the calculation of the corresponding transform of these irreducible 
segments. The theory is exact. 

The technical realization of this theory can be described as follows: 
The irreducible segments of the time evolution of the current are called ''kernels``.
We distinguish between the ''density matrix kernel`` $K$ which determines the reduced density matrix of the
quantum dot and the ''current kernel`` $K_C$ which defines the relation between the reduced density matrix
and the current. The fact that the time evolution of the current is completely determined by the kernels can 
be expressed in a compact way by the two equations: 
\begin{eqnarray*} 
   < I_l >  (t)    &=&  Tr   \int_{t_0}^t dt' K_C ( t - t' ) \rho_\odot (t'), \\
  \dot{\rho}_{\odot}(t) &=& \frac{i}{\hbar} [\rho_{\odot}(t), H_{\odot}] + \int_{t_0}^t ds K
(t-s) \rho_\odot(s),
\end{eqnarray*}
where we use the terminology $\rho_\odot (t) := Tr_R \left\lbrace \rho (t) \right\rbrace$ for the reduced 
density matrix of the 
quantum dot. The second equation is also called the quantum master equation \cite{Blum, Grifoni96, 
Weiss12}. We take the Laplace 
transform of both equations in the limit $\lambda \to 0$. Then, 
the second equation allows the calculation of 
the stationary reduced density matrix as far as $K(\lambda = 0)$ is calculated. Finally, the Laplace transform
of the first equation can be used to calculate the stationary current.

The calculation of the current means thus the calculation of the kernels. 
The contributions to the kernels are visualized by diagrams, whose number and variety is huge. 
This forces us to take into account only special 
classes of diagrams about which we have reason to believe that they might be important and which we are able
to calculate. Only in the special case of the spinless quantum dot an exact solution was presented 
\cite{Schoeller94}.

\subsection{Second order approximation}
Our approximation is an extension of the second order theory, so we recall in the following its meaning and
its main predictions.

\label{introduction of parameter w}
If we multiply the tunneling Hamiltonian by a dimension-less parameter ''$w$'', then everything becomes a 
function of $w$, including the kernels ``$K (w)$``, ``$ K_C (w) $`` and finally also the current, 
``$ < I_l >_\infty  (w) $''. All of 
the contributions
to the kernels have an order in the sense that the coefficients of the tunneling Hamiltonian appear a certain 
number of times. All odd orders vanish. Thus, they have the structure: 
\begin{eqnarray*}
 K (w) &=& w^2 K^{(2)} + w^4 K^{(4)} + \dots ,\\
 K_C (w) &=& w^2 K_C^{(2)} + w^4 K_C^{(4)} + \dots .\\
\end{eqnarray*}
For small values of $w$, i. e., for weak tunneling coupling, one  takes into 
account only the approximations for the kernels of order $2n$ and  calculates the current on this basis. One 
obtains the  current of order $2n$, ``$I^{(2n)} (w)``$, i.e., the Taylor expansion of the current around 
$w = 0$ of order $2n$. The current, Eq. (\ref{current}), is analytic in $w^2$. In the case $n= 1$ one obtains 
the second order current,
which is just of the form $w^2 \cdot constant$.

\subsubsection{Second order density matrix}
\begin{figure}[h]\centering 
\includegraphics[width = 0.4\textwidth]{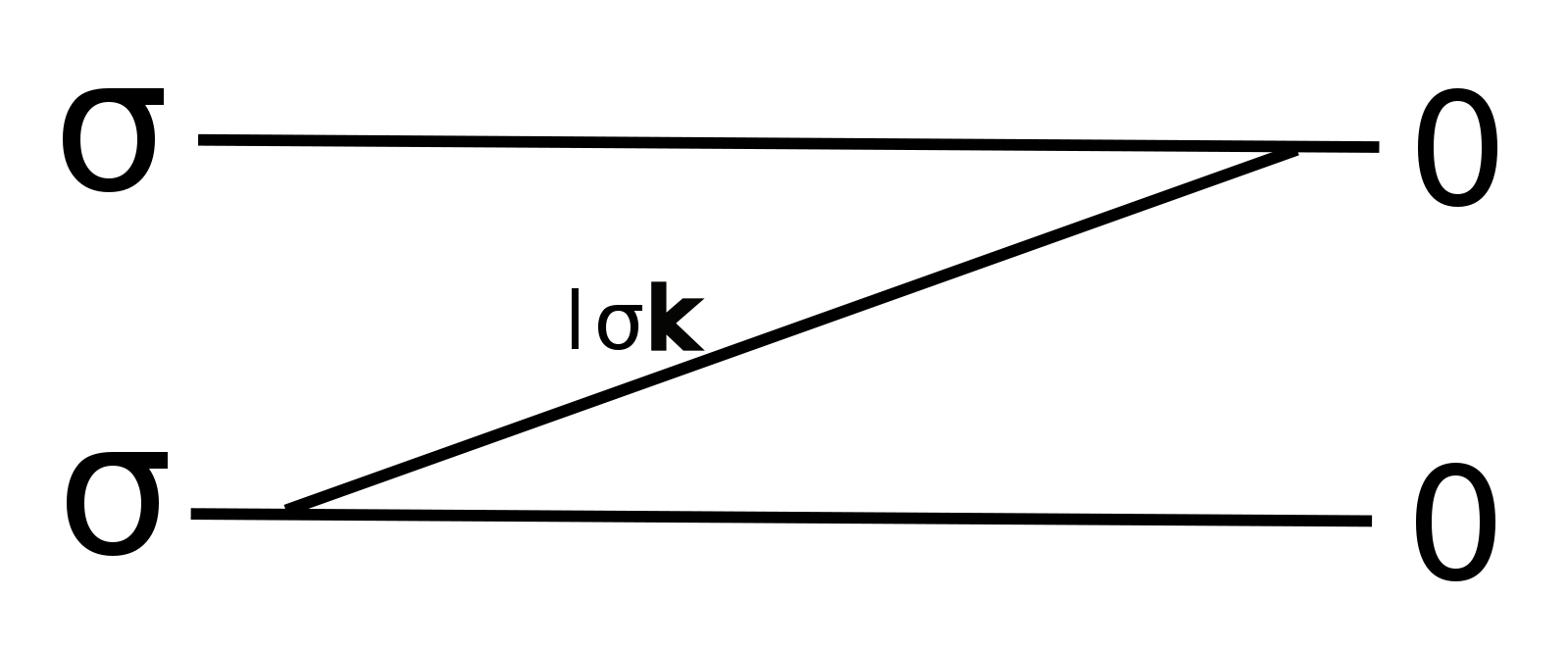}
\caption{\small  An example of a second order diagram; by convention we let the time increase from the right
to the left. The horizontal lines are called ``contours'', the third line ``tunneling line''; it represents an 
electron from lead $l$ with spin ``$\sigma$'' and wave vector ``${\bf k}$`` which tunnels in two steps onto 
the dot. The intersection points of the tunneling lines with the contours are called ''vertices``. The vertices
separate the contours into intervals, and to each of these we assign a quantum dot state. The particle 
number of neighbouring quantum dot states can differ only by $\pm 1$. The chronological order of the density 
matrices of the quantum dot in this diagram is found as follows: We imagine a vertical line which cuts each of 
the contours one time. We consider then especially the intervals between two neighbouring vertices which are 
cut by the vertical line and take the two quantum dot states assigned to them. If they are ''$a$`` and ''$b$``
on the lower and on the upper contour, respectively, then the current quantum dot matrix is given by 
$|a\rangle \langle b|$. The imaginary vertical line we move from the right to the left end of the diagram. 
The sequence of quantum dot matrices in
the above diagram is:  $|0\rangle \langle 0|$,  $ |0 \rangle \langle \sigma| $ and finally  
$|\sigma \rangle \langle  \sigma|$.  }
\label{second order}
\end{figure}

One of the diagrams visualizing the second order contributions to the density matrix kernel is shown in figure 
\ref{second order}. A possible way of describing the process is to say that the quantum dot is at first in the 
un-occupied state ``$0$''; then an electron with spin $\sigma$ tunnels in two steps onto the dot. Finally, the 
dot is in the state ``$\sigma$''. The analytical expressions which correspond to the diagrams are given by 
diagrammatic rules, e.g.  \cite{Schoeller94, König96, Koller10, Report10}. The expression for the diagram in 
Figure 
\ref{second order} reads:
 \begin{displaymath}
 \frac{1}{\hbar}  \frac{f_l (\varepsilon_{l {\bf k} \sigma}) \left| 
  T_{l {\bf k} \sigma}  \right|^2 }{ \hbar \lambda + i ( \varepsilon_{l {\bf k} \sigma} - E_{1 0}) },
\end{displaymath}
where we let $\lambda$, the argument of the Laplace transform,  still be finite. For simplicity we assume 
degeneracy, $E_\sigma = E_{\bar \sigma}$, and write $E_{\sigma 0} =: E_{10}$. Later we will consider also the 
case of different energies  $E_\sigma \neq E_{\bar \sigma}$.
We let $f_l (\varepsilon)$ be the Fermi function at chemical potential
$\mu_l$ and temperature $T$, i.e., $f_l (\varepsilon) = f \left( (\varepsilon - \mu_l)/ k_B T\right)$ with 
$f(x) = 1/(1 + e^x)$. We perform then the sum with respect to the leads and the wave vector. 
The thermodynamic limit is taken by replacing the sum with respect to 
the allowed wave vectors ${\bf k}$ by an integral over the first Brillouin zone. The expression turns into:
 \begin{displaymath}
 \frac{1}{\hbar} \sum_{l} \int d {\bf k} Z_l  \frac{f_l (\varepsilon_{l {\bf k} \sigma}) \left| 
  T_{l {\bf k} \sigma}  \right|^2 }{ \hbar \lambda + i ( \varepsilon_{l {\bf k} \sigma} - E_{1 0}) },
\end{displaymath}
where $Z_l$ is the number of allowed wave vectors in the first Brillouin zone per volume in the wave vector
space. We split now the integration into two parts: First, we fix the band energy and integrate over the 
surface in the first Brillouin zone where $\varepsilon_{l {\bf k} \sigma}$ equals this band energy. In a second
step, we integrate over the band energies \cite{Ashcroft}. The integral turns into: 
 \begin{displaymath}
  \sum_{l} \frac{Z_l}{\hbar}  \int d\varepsilon    \frac{f_l (\varepsilon)  }
{ \hbar \lambda + i ( \varepsilon - E_{1 0}) }
\int_{ \left\lbrace \varepsilon_{l {\bf k} \sigma}  = \varepsilon 
  \right\rbrace } 
  \frac {dS \left| 
  T_{l {\bf k} \sigma}  \right|^2}{ \left| \nabla \varepsilon_{l \sigma} ( {\bf k}  ) \right| }.     
\end{displaymath}

There are two diagrams of second order which are contributions to the ``kernel element'' 
$ \langle \sigma | \left \lbrace K ( \lambda ) |0\rangle \langle 0| 
\right\rbrace |\sigma \rangle $. They are given by the above diagram and by the one we get by mirroring this with 
respect to a horizontal axis. Their contributions are complex conjugate, so we have to take two times the real 
part of the above expression. In the limit $\lambda \to 0 $ we obtain: 
\begin{displaymath}
 \langle \sigma | \left\lbrace K ( \lambda = 0 ) |0\rangle \langle0| \right\rbrace |\sigma \rangle = 
\frac{2\pi}{\hbar} \sum_l 
\alpha_l^+ ( E_{1 0} ), 
\end{displaymath}
where we used the notation $\alpha_l^+ (\varepsilon) = \alpha_l (\varepsilon) f_l (\varepsilon)$ and 
\begin{equation} \label{definition of alpha}
\alpha_l ( \varepsilon ) = \int_{ \left\lbrace \varepsilon_{l {\bf k} \sigma}  = \varepsilon 
  \right\rbrace } 
  \frac {dS Z_l \left| 
  T_{l {\bf k} \sigma}  \right|^2}{ \left| \nabla \varepsilon_{l \sigma} ( {\bf k}  ) \right| }.     
\end{equation}
In the case that the tunneling coefficients $T_{l {\bf k} \sigma}$ were independent of the wave 
vector the function $\alpha_l (\varepsilon) $
would just be proportional to the density of electron levels in lead $l$. However, we point out that we do
not use 
such a simplifying assumption about the tunneling coefficients at this stage.  
The dimension of $\alpha_l$ is ``energy'', 
correspondingly the dimension of the kernel elements is ``rate''.

\begin{figure}[h]\centering 
\includegraphics[width = 0.4\textwidth]{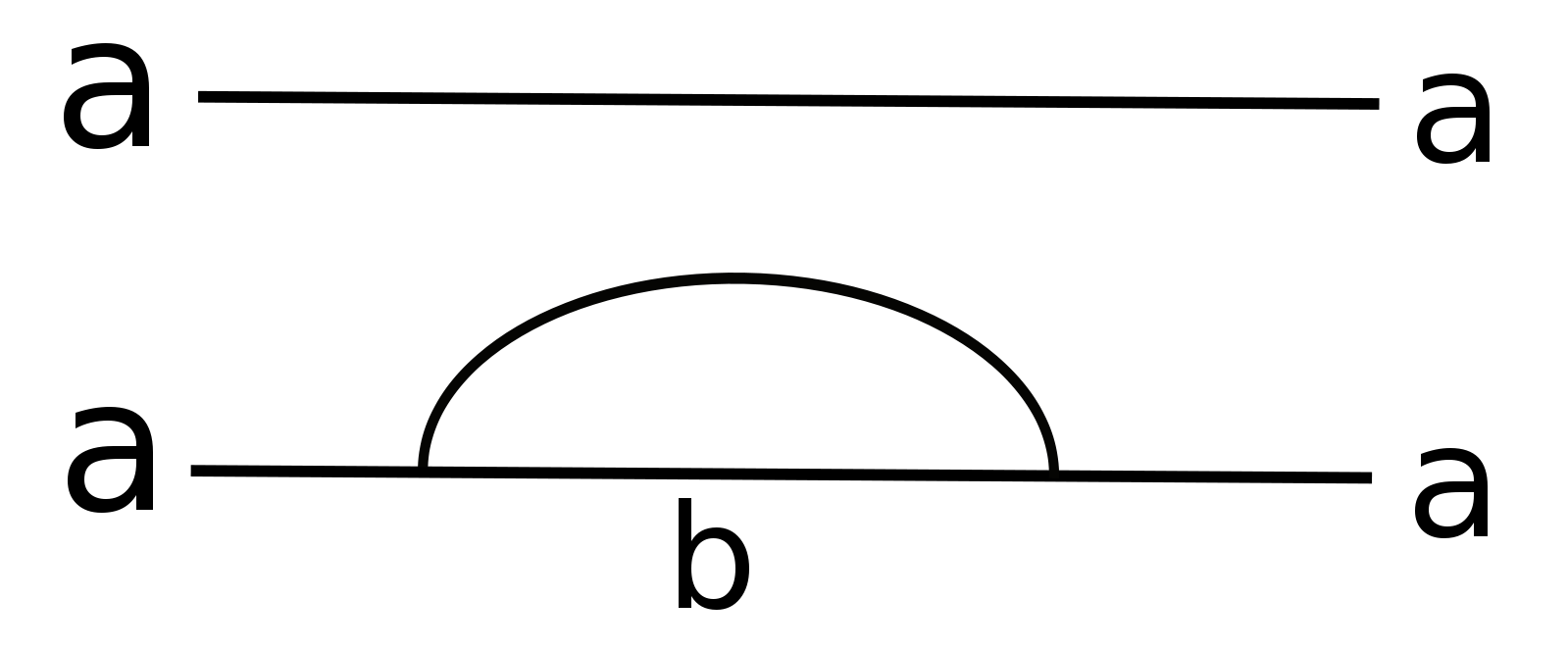}
\caption{\small  Second order diagrams of this form ensure that the trace of $K ( \lambda  ) 
|a \rangle \langle a|$ is 
always zero. Their contributions to the density matrix kernel amount to $- \sum_{ b \neq a} \langle b| 
\left\lbrace K ( \lambda ) |a\rangle    \langle a| \right\rbrace |b \rangle $.  }
\label{zero trace}
\end{figure}

The other second order kernel elements are calculated essentially  in the same way. With the further notation 
$\alpha_l^- (\varepsilon) = \alpha_l (\varepsilon) ( 1- f_l (\varepsilon))$ we obtain: 
\begin{eqnarray*}
 \langle 0 | \left\lbrace K ( \lambda = 0 ) |\sigma \rangle
   \rangle \sigma| \right\rbrace |0 \rangle  &=& \frac{2\pi}{\hbar} \sum_l
    \alpha_l^- (E_{1 0}), \\
\langle 2 | \left\lbrace K ( \lambda = 0 ) |\sigma \rangle \langle \sigma| \right\rbrace |2 \rangle  &=& 
\frac{2\pi}{\hbar} \sum_l
    \alpha_l^+ (E_{2 1 }), \\
\langle \sigma | \left\lbrace K ( \lambda = 0 ) |2 \rangle \langle 2| \right\rbrace |\sigma \rangle  
&=& \frac{2\pi}{\hbar} \sum_l
    \alpha_l^- (E_{2 1 }). \\
\end{eqnarray*}
For simplicity we assume for this a symmetry in the leads with respect to the spin, i.e., that the 
definition of $\alpha_l$,
Eq. (\ref{definition of alpha}), does not depend on the spin. By considering the contributions of diagrams as in 
figure \ref{zero trace} one can verify that the general property of the density matrix kernel $Tr \left\lbrace 
K |a\rangle \langle a| \right\rbrace = 0$ holds true also {\em within} the second order theory. Therefore, we 
already calculated  implicitly the remaining kernel elements of the form $\langle a | \left\lbrace K ( \lambda ) 
|a\rangle \langle a| \right\rbrace |a \rangle $.

For the Anderson model quantum dot, the density matrix kernel always transforms a diagonal matrix into a
diagonal matrix, i.e., $ \langle b | \left\lbrace K ( \lambda = 0 ) |a \rangle   \langle a| \right\rbrace |b' 
\rangle  = 0 $ if $ b \neq b'$.
Thus, we can say that it is a linear operator with rank three or lower acting on the four-dimensional space 
of the diagonal matrices
 since one degree of freedom is destroyed by the condition that the trace of the 
resulting matrix is zero. We can conclude that there is a diagonal solution ``$\rho$`` of the quantum master 
equation (QME) in the stationary limit. With the notations 
\begin{eqnarray*}
 \rho_{aa}          &:=&   \langle a| \rho |a \rangle , \\
\Gamma_{l, 01}^\pm  &:=&   \frac{2\pi}{\hbar} \alpha_l^\pm (E_{10}) \quad \mbox{ and }  \\
\Gamma_{l, 12}^\pm  &:=&   \frac{2\pi}{\hbar} \alpha_l^\pm (E_{21})
\end{eqnarray*}
the QME turns into the following set of two equations (for the three variables $\rho_{00}, 
\rho_{22} \mbox{ and }\rho_{\uparrow \uparrow} = \rho_{\downarrow \downarrow } $):

\begin{eqnarray*}
 0 &=& - \rho_{00} \sum_l \Gamma_{l,01}^+  +   \rho_{\sigma \sigma} \sum_l \Gamma_{l, 01}^-, \\
0  &=&  \rho_{\sigma\sigma} \sum_l \Gamma_{l,12}^+ - \rho_{22} \sum_l \Gamma_{l,12}^-.  
\end{eqnarray*}
This information is sufficient to determine the stationary reduced density matrix since we know that the 
normalization condition, $\rho_{00} + 2\rho_{\sigma\sigma} + \rho_{22} = 1$, holds. The solution is: 
\begin{equation}\label{solution_of_qme}
 \left( \begin{array}{c} \rho_{00}\\\rho_{\uparrow\uparrow}\\
  \rho_{\downarrow\downarrow}\\ \rho_{22} \end{array} \right) = 
\frac{1}{ \Gamma_{12}^- \Gamma_{01} + \Gamma_{01}^+ \Gamma_{12} } 
\left( \begin{array}{c}   \Gamma_{01}^- \Gamma_{12}^- \\  \Gamma_{01}^+ \Gamma_{12}^-   \\ 
   \Gamma_{01}^+ \Gamma_{12}^-  \\  \Gamma_{12}^+ \Gamma_{01}^+   \end{array} \right), 
\end{equation}
 where we used the notations 
\begin{eqnarray*}
  \Gamma_{ab}^\pm &:=& \sum_l \Gamma_{l,ab}^\pm ,   \\ 
    \Gamma_{ab}   &:=&  \Gamma_{ab}^+    +  \Gamma_{ab}^-.
\end{eqnarray*}

\subsubsection{Second order current kernel}
In order to calculate the current we have to determine the second order current kernel. The structure of the 
contributions to it is the same as that of the contributions to the density matrix kernel.
We take into account only the diagrams with the final vertex on the lower contour. The lead-index attached 
to the corresponding tunneling line is fixed and given by the lead onto which we are calculating the current. 
An additional sign, as compared to the density matrix kernel, has then to be taken into account. There are 
several equivalent possibilities of defining the current kernel \cite{Schoeller97}.

For example, the diagram in figure \ref{second order} yields the contribution
\begin{displaymath}
   \frac{-1}{\hbar}  \int d\varepsilon    \frac{\alpha_l^+ (\varepsilon)  }
{ \hbar \lambda + i ( \varepsilon - E_{1 0}) }
 \end{displaymath}
to the trace $Tr \left\lbrace K_C (\lambda) |0\rangle \langle 0| \right\rbrace $.

The other contribution to this trace comes from the diagram in figure \ref{zero trace} if we set 
$ a = 0, b = \sigma$. 
The two contributions are complex conjugate and so we get:  
\begin{displaymath}
 Tr \left\lbrace K_C (\lambda = 0) |0\rangle \langle 0|  \right\rbrace = -2 \Gamma_{l,01}^+. 
\end{displaymath}
In an analogous way one obtains: 
\begin{eqnarray*}
 Tr \left\lbrace K_C (\lambda = 0) |2\rangle \langle 2|  \right\rbrace &=& 2 \Gamma_{l,12}^- \quad \mbox{and} \\
Tr \left\lbrace K_C (\lambda = 0) |\sigma \rangle \langle \sigma|  \right\rbrace &=& \Gamma_{l,01}^- 
- \Gamma_{l,12}^+.
\end{eqnarray*}
The second order particle current is then found by applying the current kernel to the reduced density matrix and 
taking the trace \cite{König96}:
\begin{equation} \label{current in terms of rates}
I_l^{(2)} ( w = 1 )  =  \frac{2}{N} 
\left( \begin{array}{c} \Gamma_{12}^- \\ \Gamma_{01}^+     \\
 \end{array} \right) \left( \begin{array}{c} \Gamma_{\bar{l},01}^+ \Gamma_{l,01}
- \Gamma_{\bar{l},01} \Gamma_{l,01}^+ 
 \\ \Gamma_{\bar{l},12} \Gamma_{l,12}^- - \Gamma_{\bar{l}, 12}^- \Gamma_{l,12}     \\
 \end{array} \right),
 \end{equation}
where we used ``$\bar{l}$'' to denote the opposite lead of lead ``$l$'' and the abbreviations:
\begin{eqnarray*}
 N  &:=& \Gamma_{12}^- \Gamma_{01} + \Gamma_{01}^+ \Gamma_{12}, \\
\Gamma_{l,ab} &:=&  \Gamma_{l,ab}^+ + \Gamma_{l,ab}^-.
\end{eqnarray*}
The letter $w$ denotes the coupling parameter as introduced above; for simplicity we will leave it away in 
the following.
This is the particle current onto lead $l$. The net current, i.e., the sum of the two currents onto lead $l$ 
and $\bar{l}$, is zero. To determine the electric current one has to multiply by the electron 
charge.

In the case of proportional tunneling coupling, i.e., $\alpha_l = \kappa_l \alpha$, with $\kappa_l$  
positive scalar factors fulfilling $\sum_l \kappa_l = 1$, the expression for the current can be simplified:
\begin{eqnarray} \label{current for proportional coupling}
I_l^{(2)}  &=&  \frac{2}{1 + \frac{\Gamma_{01}^+ \Gamma_{12} } {\Gamma_{01} \Gamma_{12}^- } } 
		\left( \kappa_l \Gamma_{\bar{l}, 01}^+ - \kappa_{\bar{l}} \Gamma_{l,01}^+ \right) \nonumber \\
	   &+&  \frac{2}{1 + \frac {\Gamma_{01} \Gamma_{12}^- } {\Gamma_{01}^+ \Gamma_{12} } } 
		\left( \kappa_{\bar{l}} \Gamma_{l, 12}^- - \kappa_l \Gamma_{\bar{l},12}^- \right) .
\end{eqnarray}
The prefactor of the second line turns out to be the stationary electron number on the quantum
dot, i.e., the expectation value $Tr \left\lbrace N_\odot \rho \right\rbrace $ with $N_\odot$ 
the particle counting operator on the quantum dot;
the prefactor of the first line we might call the ``hole number'', i.e., the expectation value of 
$2 - N_\odot$.

The second order approximation can be interpreted in terms of transitions. For every pair of two quantum dot states 
``$a$'' and ``$b$'' with neighbouring particle numbers and every lead $l$ we determine a rate of transitions
``$a \to b$'' caused by the tunneling of an electron from lead $l$ onto the dot or from the dot onto lead $l$, 
provided that the dot is in the state $|a\rangle \langle a|$. The stationary density matrix is determined 
by the demand that all of these transitions compensate each other. Then we calculate the current by 
balancing the transitions. Our non-perturbative approximation is an extension of the second order theory. The 
equations for the density matrix and the current in terms of the transition rates $\Gamma_{l,ab}^\pm$ still hold 
true but the expressions for these rates change.

\section{Dressed second order diagrams}
\begin{figure}[h]\centering 
\includegraphics[width = 0.4\textwidth]{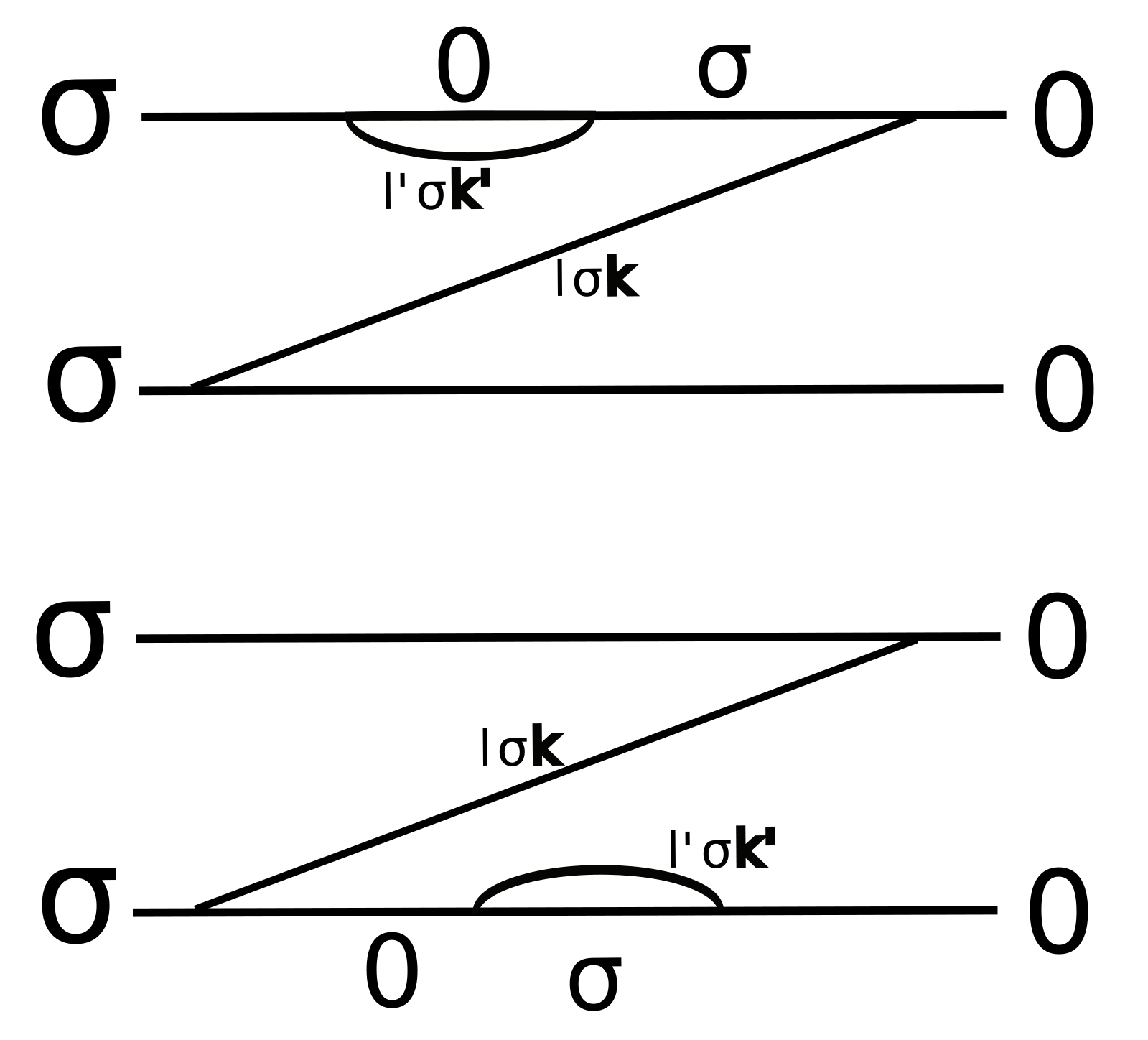}
\caption{\small  Two examples of dressing the diagram of figure \ref{second order} with further 
tunneling lines. In the upper diagram the temporal sequence of quantum dot matrices is:  $|0\rangle \langle 0|, 
|0\rangle \langle \sigma|, |0\rangle \langle 0 | , |0\rangle \langle 
\sigma |$ and $| \sigma \rangle \langle \sigma |$. A hole from lead $l'$ with wave vector ${\bf k'}$ is 
participating in the process. In the lower diagram, it is a further {\em electron} which is accompanying the 
tunneling of the electron from the level $l {\bf k} \sigma$. }
\label{dressed second order example}
\end{figure}

In this section we account for diagrams similar to the ones of the second order theory but ``dress`` them by charge fluctuations.
Figure \ref{dressed second order example} shows two possibilities of dressing the second order 
diagram of figure \ref{second order}. Apart from the electron that tunnels in two steps onto the dot there 
is one more electron level of the leads involved. We might say that an electron
(lower example diagram) or a hole (upper example diagram) tunnels for some time halfway onto the dot and then 
leaves it again. For linguistic simplicity we restrict ourselves here to speaking about particles tunneling 
{\em onto} the 
dot, with the consequence that we are using the terms ''hole`` as well as ''electron``. 
The tunneling of the one electron which finally enters the dot is accompanied 
by the  tunneling of further electrons or holes in these diagrams. 

According to the diagrammatic rules the sum of the contributions of the diagrams in figure 
\ref{dressed second order example} to the density matrix kernel is given by
\begin{eqnarray*}
&& \frac{1}{\hbar} \int d\varepsilon \frac{ \alpha^+ (\varepsilon) } {\hbar \lambda + i(\varepsilon - E_{10} ) } \\
&& \frac{ -1 }  { \hbar \lambda + i (\varepsilon - E_{10})  } \int d\varepsilon'
 \frac{ \alpha (\varepsilon')  } { \hbar \lambda + i (\varepsilon - \varepsilon') },
\end{eqnarray*}
where we let $\lambda$ still be finite and used the notations:
\begin{eqnarray*}
 \alpha^\pm &:=& \sum_l \alpha_l^\pm , \\
 \alpha     &:=& \alpha^+ + \alpha^- = \sum_l \alpha_l.
\end{eqnarray*}
In the first line we recognize the contribution of the second order diagram. However, the integrand is multiplied
by a factor, the second line, and this reflects the participation of further particles. From the upper diagram in 
figure  \ref{dressed second order example} we get ''$\alpha^-(\varepsilon')$``, from the lower one we get 
''$\alpha^+(\varepsilon')$``. The sum yields ''$\alpha(\varepsilon')$`` which appears in the factor.

\begin{figure}[h]\centering 
\includegraphics[width = 0.4\textwidth]{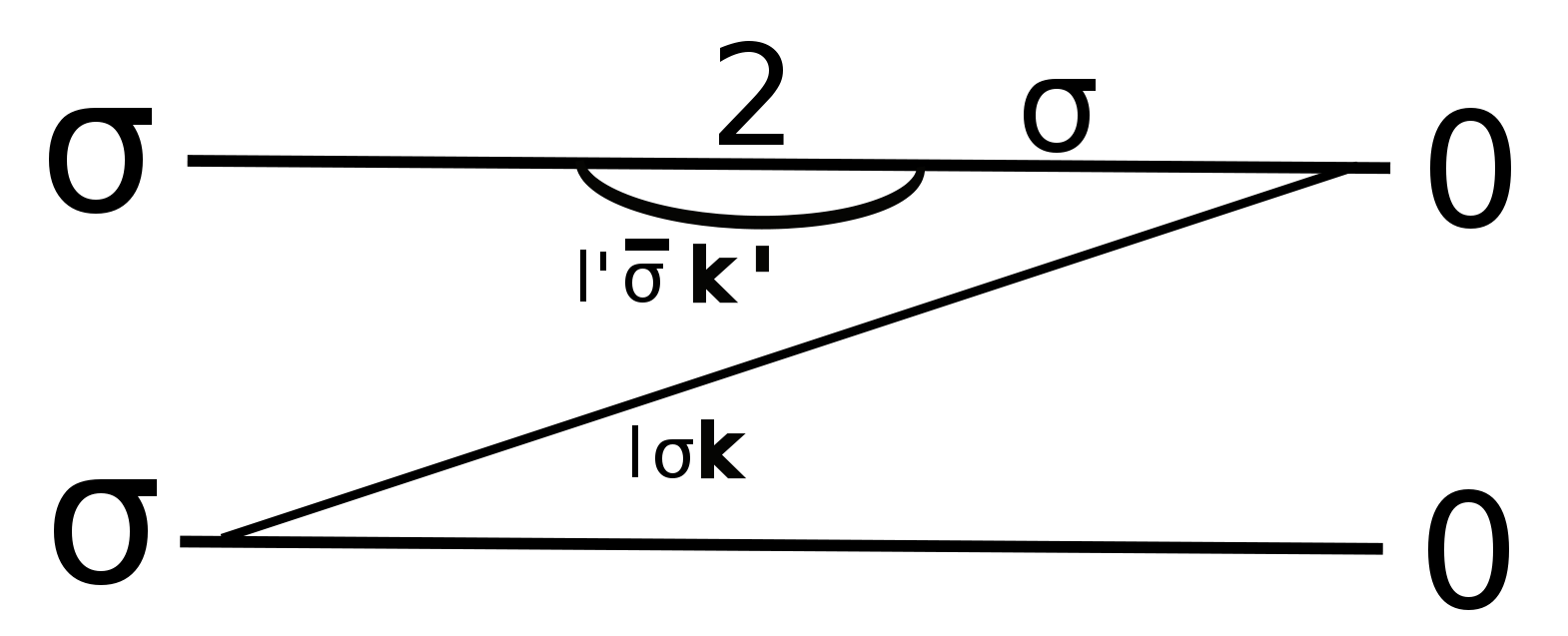}
\caption{\small The existence of the state ''2'' leads to a fourth possibility of dressing. The tunneling line
on the upper contour represents an electron of the opposite spin which tunnels onto the dot and leaves it again.}
\label{terrible diagram}
\end{figure}

Because of the existence of the two spins there is still a third way of dressing our second order diagram with 
one further
tunneling line: The bubble on the lower contour of figure \ref{dressed second order example} might as well 
represent an electron with {\em opposite } spin which accompanies the tunneling. The contribution of this diagram
is the same as the contribution of the diagram with only one spin appearing, but it is important because it 
does not have a counterpart: There is {\em no way } of dressing the diagram with a bubble on the upper 
contour which represents a hole of the {\em opposite} spin. Finally, there is the possibility to dress the diagram
by a tunneling line on the upper contour which represents an electron of the opposite spin, as shown in figure
\ref{terrible diagram}.

\begin{figure}[h]\centering 
\includegraphics[width = 0.5\textwidth]{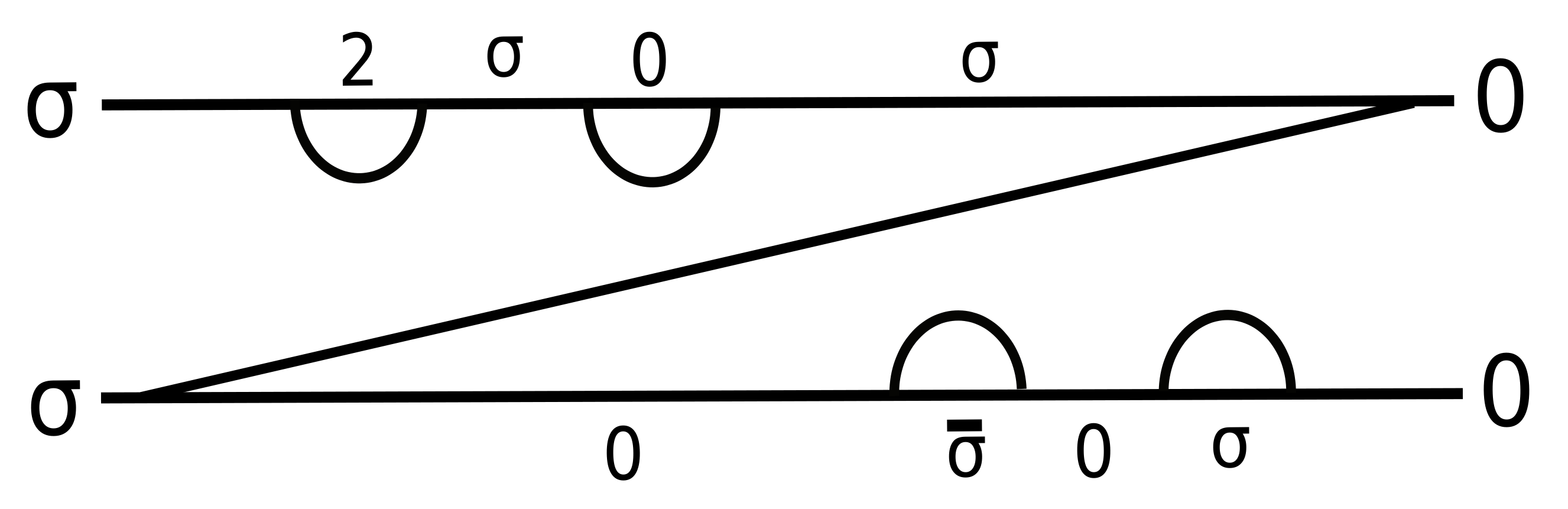}
\caption{\small  An example of a diagram with four ''bubbles'' and one ``long'' tunneling line. If we count the 
bubbles from the right to the left, then for the choice of each of them we have essentially four different 
possibilities. This example-diagram contains all four of these possibilities.}
\label{more bubbles}
\end{figure}
We saw that there are essentially four ways of dressing the second order diagram with one bubble. Moreover, 
we can dress the diagrams with two or even more, in general: $n$, subsequent, non-intersecting, bubbles. 
An example is sketched 
in figure \ref{more bubbles}. For the choice of each of these bubbles we have four possibilities. It might 
represent  an electron of the same or the opposite spin and thus appear on the lower contour or represent a 
hole of the same spin or an electron of the opposite spin and appear on the upper contour. The sum of the  
contributions of all of these diagrams to the kernel element $\langle \sigma | \left\lbrace K (\lambda) |0\rangle 
\langle 0|  \right\rbrace | \sigma \rangle$ is
\begin{eqnarray*}
 && \frac{1}{\hbar} \int d\varepsilon \frac{ \alpha^+ (\varepsilon) } {\eta + i(\varepsilon - E_{10} ) } 
 \sum_{n=0}^\infty  \left\lbrace \frac{ -1 }  { \eta + i (\varepsilon - E_{10})  }  \right\rbrace^n \\
&& \left\lbrace \int d\varepsilon'\frac{ (\alpha + \alpha^+) (\varepsilon')  } { \eta + i 
(\varepsilon - \varepsilon') }
+ \frac{\alpha^+ (\varepsilon')}  { \eta + i ( \varepsilon + \varepsilon' - E_{20} ) }
 \right\rbrace^n  = 
\end{eqnarray*}
\begin{displaymath}
   \int d\varepsilon \frac{\alpha^+(\varepsilon) / \hbar} { \eta + i(\varepsilon - E_{10})
+ \int d\varepsilon' \frac{ (\alpha + \alpha^+) (\varepsilon') } { \eta + i(\varepsilon - \varepsilon') 
 } + \frac{\alpha^+ (\varepsilon')}  { \eta + i ( \varepsilon + \varepsilon' - E_{20} ) } },
\end{displaymath}
where we replaced ``$\hbar \lambda$'' by ``$\eta$''. 
This summation of a geometric series can be justified for ''large enough`` values of $\lambda$. However,
we are interested in the limit $\lambda \to 0$. For this we remember that we want to calculate the 
Laplace-transform of the sum of the corresponding diagrams in the {\em time space} and let $\lambda$ be complex.
If the real part of $\lambda$ is sufficiently large, say, $Re \lambda > c$, then the Laplace transform of the
sum of the diagrams in the time space is indeed given by the right hand side of the above equation.  
One can represent now the sum of the diagrams in the time space as the Laplace back transform of this right
hand side\cite{Laplace transform}
and in this way see that its Laplace transform exists for all values of $\lambda$ with strictly
positive real part. The right hand side of the above equation, {\em too}, is holomorphic in $\lambda$ on the whole
half plane $\left\lbrace Re > 0 \right\rbrace $ where the real part is positive. Thus, we have {\em two} 
holomorphic functions on $\left\lbrace Re > 0 \right\rbrace$ which are equal on 
$\left\lbrace Re > c \right\rbrace$.
The theory of holomorphic functions says that they must be equal everywhere.

\subsection{DSO tunneling rates}
All of the other second order diagrams can be dressed in the same way. For the diagrams connecting the particle
numbers one and two we see that a support of the tunneling by holes and electrons of the same spin and by 
a hole of the opposite spin is possible, but not by an electron of the opposite spin. We obtain the 
following transition rates within this dressed second order approximation: 
\begin{widetext}
\begin{eqnarray} \label{DSO transition rates}
\Gamma_{l,01}^\pm &=& \frac{2\pi}{\hbar} \int d \varepsilon \frac { \alpha_l^\pm (\varepsilon) 
\left[ (\alpha + \alpha^+)(\varepsilon) + \alpha^+ ( E_{20} - \varepsilon )    \right] }
 { \pi^2 \left[ (\alpha + \alpha^+) (\varepsilon) + \alpha^+(E_{20} - \varepsilon) \right]^2  + 
\left[ \varepsilon +  p_{\alpha + \alpha^+} (\varepsilon) - p_{\alpha^+} (E_{20} - \varepsilon)  - E_{10}
\right]^2  },  \\ \label{DSO transition rates 2}
\Gamma_{l,12}^\pm &=& \frac{2\pi}{\hbar} \int d \varepsilon \frac { \alpha_l^\pm (\varepsilon) 
\left[ (\alpha + \alpha^-)(\varepsilon) + \alpha^- ( E_{20} - \varepsilon )    \right] }
 { \pi^2 \left[ (\alpha + \alpha^-) (\varepsilon) + \alpha^-(E_{20} - \varepsilon) \right]^2  + 
\left[ \varepsilon +  p_{\alpha + \alpha^-} (\varepsilon) - p_{\alpha^-} (E_{20} - \varepsilon)  - E_{21}
\right]^2  },
\end{eqnarray}
\end{widetext}
where we define for any function $h$ the function $p_h$ by 
\begin{equation} \label{principal part}
 p_h (\varepsilon) := \int d\omega \frac{h ( \varepsilon + \omega ) - h (\varepsilon - \omega)}{\omega}. 
\end{equation}

Hence, the DSO rates are given in the form of an integral where the integrand is the product of the second order
functions $\alpha_{l}^\pm (\varepsilon) $ and of a Lorentzian-like resonance function. We thus expect that 
the second order rates are recovered when the temperature broadening of the functions $\alpha_l^\pm $ largely 
exceeds the width of the Lorentzian broadening.  
In the limit of weak coupling these transition rates indeed turn into the transition rates of the second order 
theory. More precisely, we get if we multiply the tunneling Hamiltonian with a parameter $w$ as done in 
\ref{introduction of parameter w}:
\begin{displaymath}
\frac{\Gamma_{l,ab}^\pm ( \mbox{dressed second order} ) (w) }{ \Gamma_{l,ab}^\pm 
(\mbox{second order}) (w) } \to 1 (w\to 0). 
\end{displaymath}

The stationary reduced density matrix within the DSO is given by Eq. (\ref{solution_of_qme}) and the current is 
given by Eq. (\ref{current in terms of rates}) and in the case of proportional coupling by 
Eq. (\ref{current for proportional coupling}). 

In Ref. \cite{König96}  a diagram selection called the 
''resonant tunneling approximation`` (RTA) was applied: The case of 
infinite interaction was considered and therefore the diagrams containing the state ''two'' do not contribute; 
transition rates were derived, the 
stationary density matrix was determined and finally the current was obtained. 
The RTA takes into account all of the diagrams within the DSO which do not contain the state "2``. However, 
there are many diagrams outside the DSO which are contained in the RTA as we will discuss in Sec. 
\ref{spinless quantum dot}.
Comparisons between the predictions of the DSO and RTA will be performed in the cases of the spinless 
quantum dot and of the SIAM with infinite interaction.

\subsection{Linear conductance within the DSO}
We assume that we can obtain the second order functions $\alpha_l$  by multiplication of the density of electron 
levels in the leads by a coupling constant while in general their definition is more complicated 
(Eq. (\ref{definition of alpha})).
About the density of electron levels we make simplifying assumptions such that 
we can concentrate on effects which are not due to special behaviour of the density of electron levels. 
In particular, we place the Fermi level in a point with respect to which the density of electron levels is 
symmetric. Then the chemical potential at equilibrium coincides with the Fermi level for all temperatures,
see figure \ref{figure explaining our choice of b}.
By deriving the formula for the current, Eq. (\ref{current for proportional coupling}), 
with respect to the bias, $e V_{bias} = \mu_{l} - \mu_{{\bar l}}$ , one obtains at zero bias the following 
expression for the linear conductance:
\begin{widetext} 
\begin{equation} \label{linear conductance}
 G^{(DSO)} = 4 \kappa_l \kappa_{\bar l} \frac{e^2}{h}   
    \left( \begin{array}{c} \left. \left( 2 - n_\odot \right) \right|^{V_{bias} = 0} \\  
            \left.  n_\odot  \right|^{V_{bias} = 0}     \\
 \end{array} \right)
\left( \begin{array}{c}  
\int  d\varepsilon  \left. \frac { \pi^2  \alpha (\varepsilon) \left[ (\alpha + \alpha^+) (\varepsilon)
	+ \alpha^+ ( E_{20} - \varepsilon ) \right] }
	{ d_{01} (\varepsilon) } \right|^{V_{bias} = 0}  
       \frac{-1}{k_B T} f' \left( \frac{ \varepsilon - E_F  }
      { k_B T  }    \right) 
\\  
 \int  d\varepsilon  \left. \frac { \pi^2  \alpha (\varepsilon) \left[ (\alpha + \alpha^-) (\varepsilon)
	+ \alpha^- ( E_{20} - \varepsilon ) \right] }
	{ d_{12} (\varepsilon) } \right|^{V_{bias} = 0}  
       \frac{-1}{k_B T} f' \left( \frac{ \varepsilon - E_F  }
      { k_B T  }    \right)
               \\
 \end{array} \right),
\end{equation}
\end{widetext}
where $E_F$ is the Fermi level, $f(x) = 1/(1 + e^x)$ the normalized Fermi function, 
\begin{displaymath}
 n_\odot = \frac{2}{1 + \frac{ \Gamma_{01} \Gamma_{12}^-  }{ \Gamma_{01}^+ \Gamma_{12}  }   }
\end{displaymath}
is the particle number on the dot as noted above, and where we use the abbreviations ''$d_{01} (\varepsilon)$`` 
and ''$d_{12} (\varepsilon)$`` for the denominators in the expressions for the transition rates, Eqs.  
(\ref{DSO transition rates}) and  (\ref{DSO transition rates 2}), respectively.  
The prefactor $4 \kappa_l \kappa_{\bar l}$  is one in the case 
of symmetric coupling and less than one otherwise. We finally included the electron charge into the formula.

\section{Result of the DSO for a spinless quantum dot}
We consider here the case of a spinless quantum dot with only two possible states 
''$0$'' and ''$\sigma$'', so with only one spin. One obtains it by not performing the sum with respect to the 
spin in the Hamiltonian. 
This problem is equivalent to the SIAM with $E_\uparrow = 
E_\downarrow$ and 
$U = 0$ in the sense
that current across the SIAM quantum dot is then just two times the current across a spinless quantum dot.  

\label{spinless quantum dot}

The contribution of all diagrams outside the RTA 
to the kernels is zero \cite{König95} for the spinless quantum dot and thus, the RTA is exact in this case. 
The relation between our approximation and 
the diagrams of the RTA is described by figure \ref{resonant tunneling}. 
In the case of the spinless quantum dot treated here as well as in the case of infinite interaction the figure
characterizes the relation between the diagram selections completely.

The DSO approximation  
concentrates on the second order diagrams dressed by further tunneling lines, the RTA takes these ones in order 
to construct even more diagrams. The resulting diagrams 
are a combination of an integer number of DSO-diagrams. The transition rates of the DSO in the case of the 
spinless quantum dot we get by taking into account only those diagrams within the DSO which contain only the 
quantum dot states ``$0$'' and ``$\sigma$'':
\begin{figure}[h]\centering 
\includegraphics[width = 0.5\textwidth]{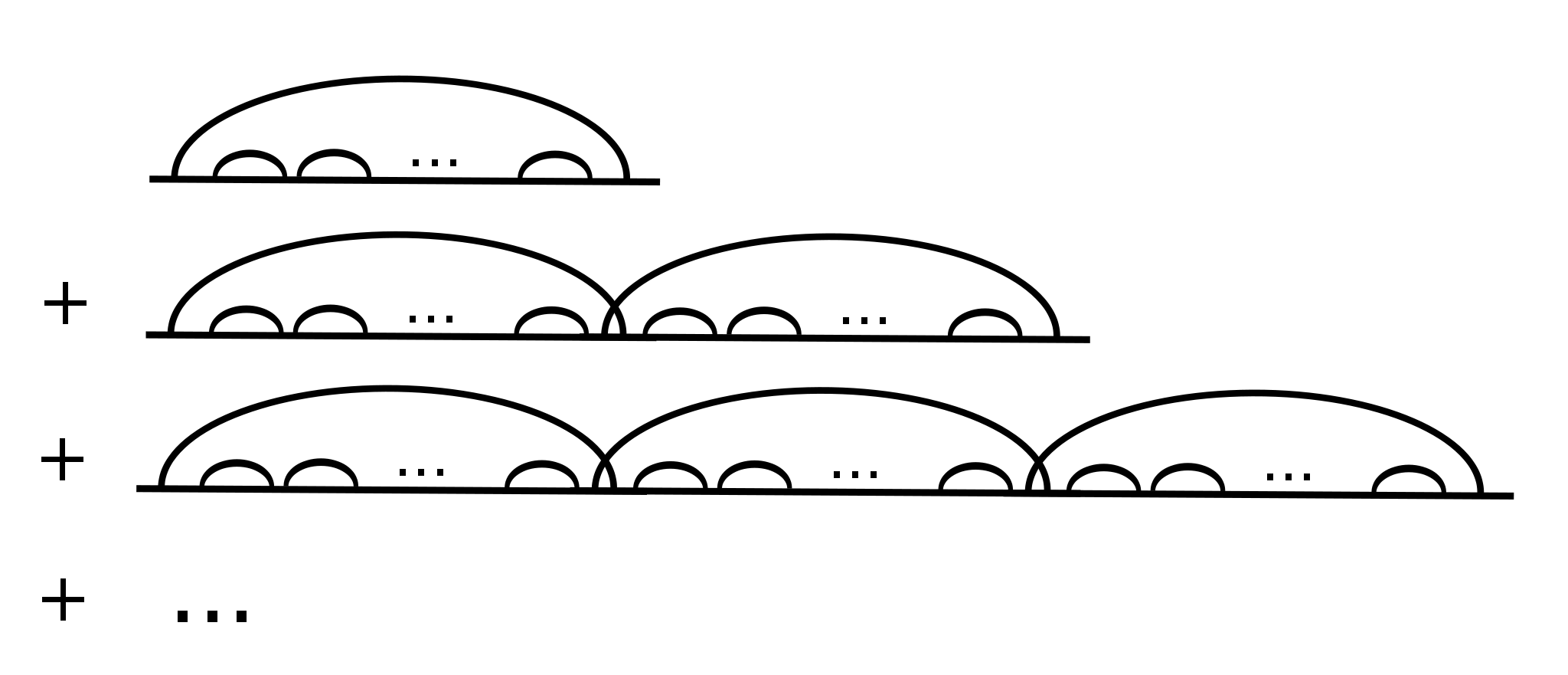}
\caption{\small  The structure of the diagrams within the RTA. When projecting 
all of the tunneling lines onto one contour only, the first line shows the diagrams within the DSO; in the 
following lines, two, three, in general: $n$ diagrams of this kind are combined to 
create new ones out of them. For the sum of all of these diagrams, an integral equation was 
derived \cite{Schoeller94}. Its origin is the relation between the contributions of all combinations of $n$ or
less diagrams of the DSO on the one hand and of all combinations of $n+1$ or 
less DSO-diagrams on the other hand. In the limit $n\to \infty$ the two are equal. }
\label{resonant tunneling}
\end{figure}
\begin{displaymath}
 \Gamma_{l,01}^\pm = \frac{2\pi}{\hbar} \int d \varepsilon \frac { (\alpha_l^\pm \alpha)(\varepsilon)}
 { \pi^2 \alpha^2 (\varepsilon) + \left( p_{\alpha} (\varepsilon) + \varepsilon - E_{10}
\right)^2  }.
\end{displaymath}
We get the density matrix
\begin{displaymath}
  \left( \begin{array}{c} \rho_{00}\\\rho_{11}\\ \end{array} \right) = \frac{1}{\Gamma_{01}^+ + \Gamma_{01}^-}
  \left( \begin{array}{c} \Gamma_{01}^-\\\Gamma_{01}^+\\ \end{array} \right)
\end{displaymath}
and the particle current in the case of proportional coupling:
\begin{displaymath}
 I_l^{DSO} = 4 \kappa_l \kappa_{\bar l}  \frac { \pi^2 } { h } \int d\varepsilon 
\frac{ \alpha^2 (\varepsilon) \left( f_{\bar l} - f_l \right) (\varepsilon)  }  
{ \pi^2 \alpha^2 (\varepsilon) + \left( \varepsilon + p_\alpha (\varepsilon) - E_{10} \right)^2  }.
\end{displaymath}
In the limit of small temperatures and in case the second order functions $\alpha_l (\varepsilon) $ are 
rather constant the current is obtained by integrating a Lorentzian-like function with width (full width at 
half maximum) $\Gamma := 2 \pi \alpha $ between the two chemical potentials. The differential conductance 
as function of the bias thus reproduces the shape of this Lorentzian. Frequently, the quantity $\Gamma $ 
rather than $\alpha$ is used to define the coupling.

In the case of proportional tunneling coupling the result of the DSO for the spinless quantum dot is actually 
the {\em same} which was presented within the RTA \cite{Schoeller94} and thus exact.
In the case of non-proportional tunneling coupling the results become different.

The diagram selection defined by the first line in figure \ref{resonant tunneling} we might call the ``simple''
selection since the pair formation of $2n$ subsequent times is one of the simplest possible irreducible pair 
formations. 
In the case of finite interaction the DSO diagram selection contains {\em less} diagrams than the simple selection.
The DSO for finite $U$ does not describe the noninteracting limit, where $U=0$, correctly. This can be seen 
by comparing the formulas for the linear conductances of the DSO in case $U=0$ on the one hand and of the DSO
in the spinless case on the other hand. However, we know by now that the {\em simple} diagram selection 
{\em does} describe the noninteracting limit correctly. Moreover, the simple selection seems to suggest the 
natural way to extend the RTA to the case of finite $U$, even though the diagram summation might be 
technically difficult. The simple selection we want to discuss elsewhere.
We will now concentrate on applying the DSO approximation to cases with nonzero interaction. We will consider
the regimes $\Gamma \sim k_B T  $ and $\Gamma \gg k_B T $ and ask with respect to which aspects the DSO 
is successful in explaining experimental results and how it compares with existing theories.   


\section{The case of infinite interaction}


\begin{figure}[h]\centering 
\includegraphics[width = 0.5\textwidth]{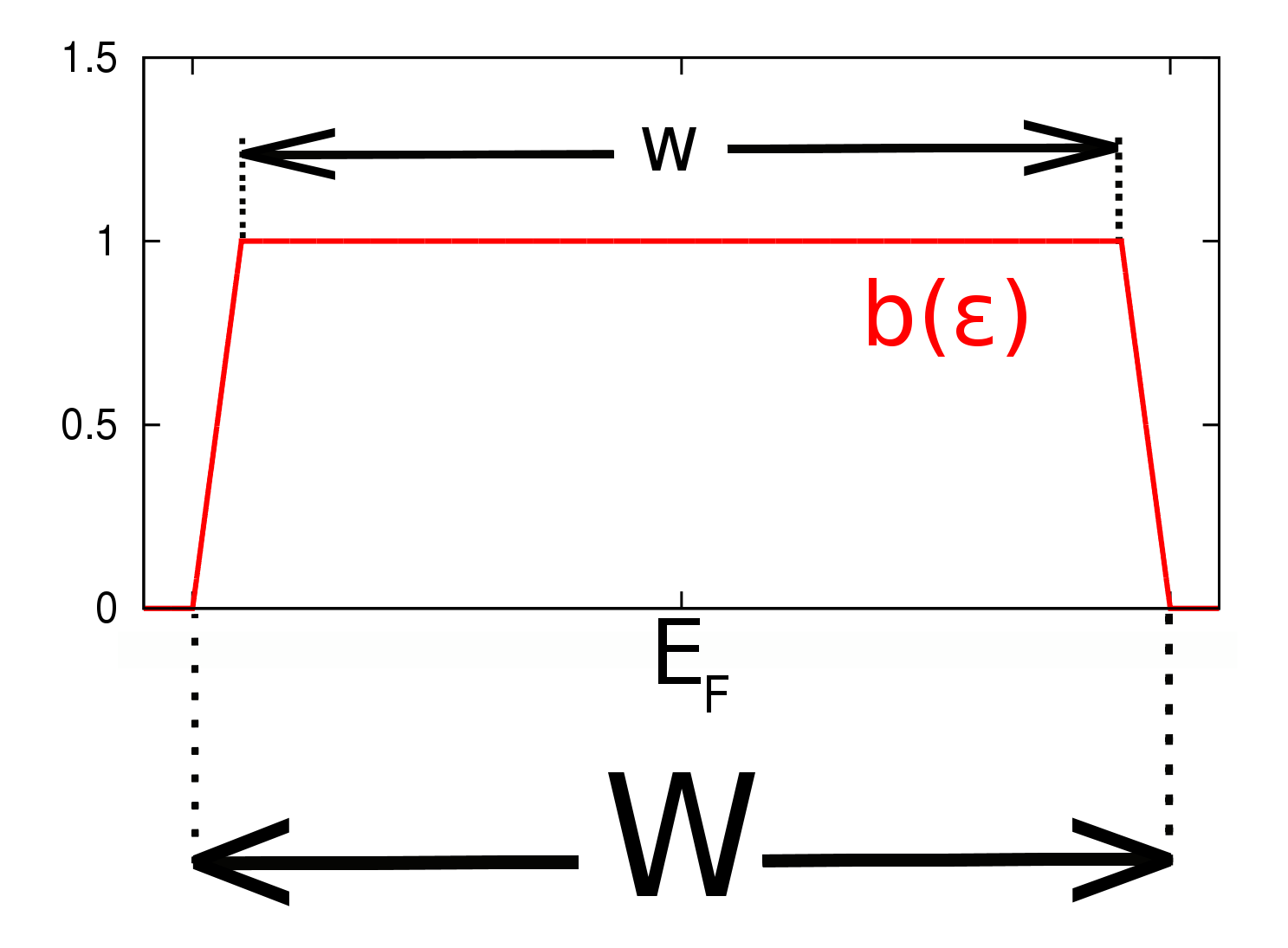}
\caption{\small Energy dependence of the dimensionless function $b(\varepsilon) = \alpha(\varepsilon) / 
\alpha (E_F)$. We placed the Fermi level $E_F$ in the middle in order
to ensure that the chemical potential at equilibrium always equals the Fermi level. Moreover, we chose 
$ W = 1eV, w = 0.9 W$. A cut-off is needed in order to ensure the existence of the principal parts $p_\alpha,
p_{\alpha^+}$ etc. of Eq. (\ref{principal part}).}
\label{figure explaining our choice of b}
\end{figure}

In the case of infinite interaction, $U = \infty$, it makes sense to neglect all of the diagrams which contain 
the state ''$2$``, assuming $\rho_{00} + \sum_\sigma \rho_{\sigma \sigma} = 1$. This was done within the RTA
and a result for the current was derived \cite{König96}. One can do the same with the DSO.  
 The formulas for the linear conductance of the RTA and DSO in the ''$U = \infty$ case`` read:
\begin{eqnarray*}
 G^{RTA}  &=&  4 \kappa_l \kappa_{\bar l} \frac{e^2}{h}   2 \\
          & &    \int  d\varepsilon  \frac { \pi^2 \alpha^2  (\varepsilon)  }
	{ d (\varepsilon)  }   
          \frac{-1}{k_B T} f' \left( \frac{ \varepsilon - E_F  }
      { k_B T  }    \right) ,   \\
G^{DSO}  &=&  4 \kappa_l \kappa_{\bar l} \frac{e^2}{h}   \left( 2 - n_\odot \right)   \\
          & &    \int  d\varepsilon  \frac { \pi^2 \left[\alpha (\alpha + \alpha^+)\right] (\varepsilon)  }
	{ d (\varepsilon)  }   
          \frac{-1}{k_B T} f' \left( \frac{ \varepsilon - E_F  }
      { k_B T  }    \right) ,
\end{eqnarray*}
 where we used the abbreviation $d( \varepsilon) := \pi^2 (\alpha + \alpha^+)^2 (\varepsilon) + 
( \varepsilon + p_{\alpha + \alpha^+} (\varepsilon) - E_{10} )^2$ for the common denominator. 
Differences are only found in the prefactor and in the numerator. To compute the conductances we have to make 
a choice about the second order function $\alpha (\varepsilon)$. We wrote $\alpha (\varepsilon) = \alpha (E_F) 
\cdot b (\varepsilon)$ with a dimensionless function $b (\varepsilon)$ fulfilling $b(E_F) = 1$. The variable 
$\alpha (E_F)$ is then our coupling parameter. Figure \ref{figure explaining our choice of b} shows how we chose 
the function $b (\varepsilon)$.

\subsection{Coulomb peaks from high to low temperatures}
\begin{figure}[h]\centering 
\includegraphics[width = 0.5\textwidth]{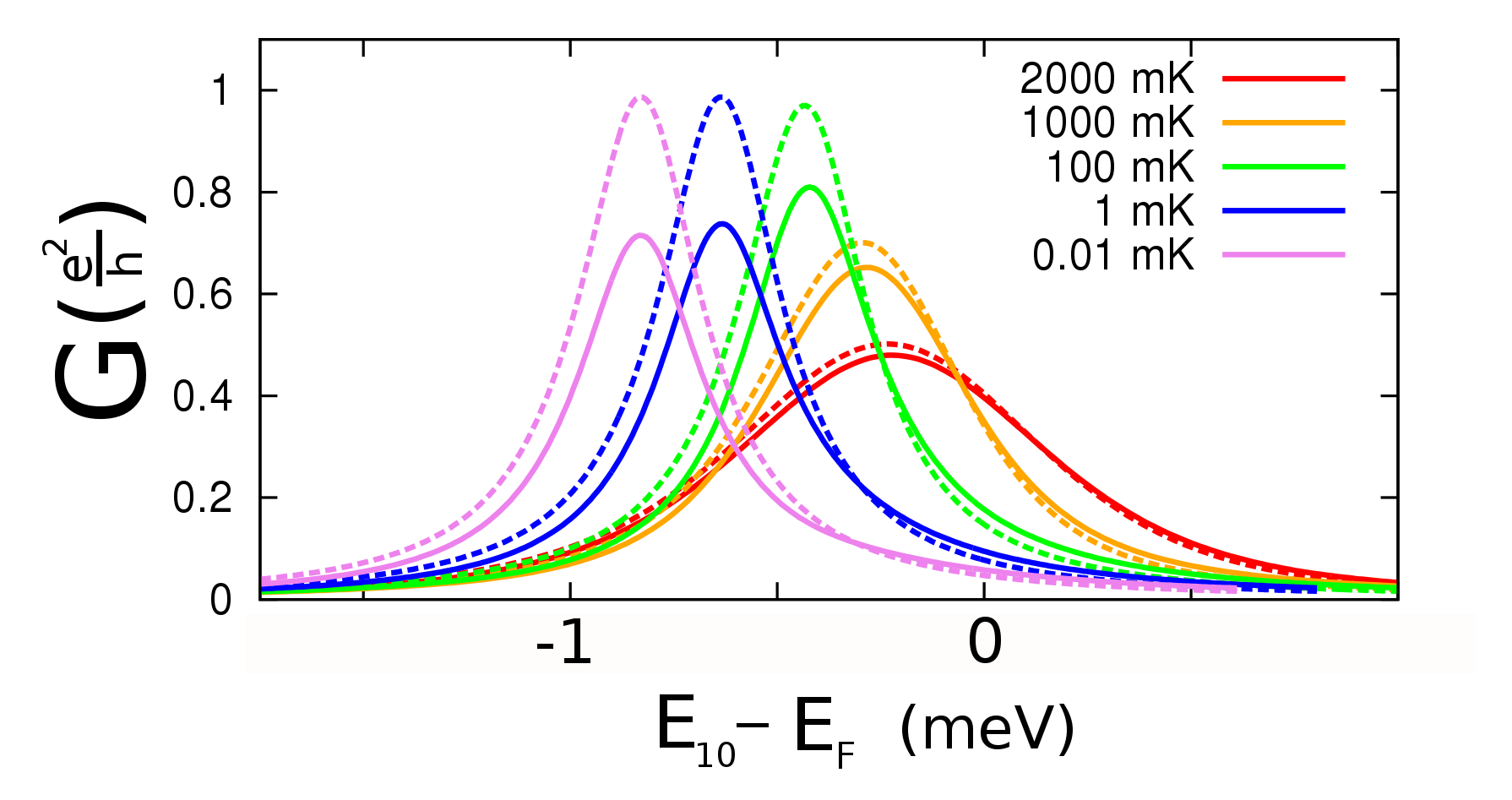}
\caption{\small (Colour online.) Linear conductance of the DSO as a function of the energy difference $E_{10}$ for different 
temperatures. The dashed lines show the result of the RTA; the coupling we chose to be $\alpha (E_F) = 0.042meV  $
and $W = 1 eV$. (The same choices we made for a later comparison with an experiment, discussed in Sec. 
\ref{later comparison}.) 
A temperature of $1K$ corresponds to a thermal energy of about $k_B T \approx 0.1 meV$. 
For large temperatures the 
resonance is smeared out, its centre is found 
roughly around the Fermi level. For decreased temperatures the position of the maximum is shifted and 
the width of the peak is proportional to $\alpha (E_F) $. While the logarithmic shift of the peak with the 
temperature does not stop, the shape of the curve and its maximum value saturate. We find numerically 
within the RTA the maximum $\approx 0.99 e^2/h$, within the DSO $\approx 0.69 e^2/h$. We 
emphasize that neither these maximum values nor the shapes of the curves in the limit of small temperatures 
depend on the the way in which the function $b(\varepsilon) $ is chosen.     
 }
\label{G_RTA_and_DSO}
\end{figure}

In figure \ref{G_RTA_and_DSO} we compare the linear conductances as a function of 
the gate voltage obtained within the RTA and the DSO for various temperatures. 
We observe a transition from a temperature dominated to a tunneling dominated width of the Coulomb peak. The 
transition occurs at temperatures around $1K$ which corresponds to a thermal energy which is of the order of 
the chosen coupling $\alpha(E_F) = 0.042 meV$. The peak height of the DSO still increases  up to temperatures 
of about $100 mK$ and decreases then. In this respect, the DSO fails to describe experimental reality below 
$100 mK$. 
Notice that the shape of the curve 
saturates at low temperatures both within the RTA and the DSO: the effect of decreasing the temeperature further 
and further is only a shift of the graph. As we will show in the next subsection, these features do not depend on 
the choice of $b(\varepsilon)$.


\subsection{Universality and Kondo temperature in the infinite $U$-case}
We show now that the DSO conductance displays universality as function of the temperature in the regime of 
strong coupling. 
For a fixed value of the gate voltage, i.e., for fixed $E_{10}$, the linear conductance becomes a function 
''$G(T)$'' of the temperature. This is expected to display universality \cite{Costi94} in the following sense: 
There is a temperature $T_K$ such that $G( T ) / G_{max} $ is a universal function of the ratio $T/T_K$, 
where $G_{max}$ is the maximum value of the conductance. This statement we can show about $G^{DSO}$ as well 
as about $G^{RTA}$. Both the RTA and the DSO, however, do not yield the expected convergence of 
$G (T) \rightarrow G_{max} = 4 \kappa_l \kappa_{\bar l} \cdot 2 e^2/h \quad ( T \to 0)$.

As we will show in the appendix one can represent $G^{DSO}$ in the form
\begin{displaymath}
 G^{DSO} \approx  4 \kappa_l \kappa_{\bar l} \frac{e^2}{h}   \left( 2 - n_\odot \right)   
F^{DSO} \left( \frac{E_{10} -             
{\bar E_{10}} }{\alpha(E_F)}, \frac{k_BT}{\alpha(E_F)} \right),
\end{displaymath}
where the definition of $F^{DSO}$ is 
\begin{displaymath}
 F^{DSO} (a,b) = \int dx \frac { -\pi^2 f'(x) (1+f(x))  }  { \pi^2 ( 1 + f(x))^2 + 
\phi_{a,b}^2 (x)  }
\end{displaymath}
and where 
\begin{displaymath}
{\bar E_{10}} = E_F + \alpha ( E_F )   p_{b^+_{T_{\alpha (E_F)}}} ( E_F ) .
\end{displaymath}
Finally, the function $\phi_{a,b} (x) =   g(x) - a + xb +  log(b) $ is defined by the use of Eq. 
(\ref{definition of g}). The temperature $ T_{\alpha (E_F)} $  we define by the demand  
$k_B T_{\alpha (E_F)} = \alpha (E_F)$. The function $b^+_T ( \varepsilon ) $ is given by
$ b^+_T (\varepsilon) = b  ( \varepsilon ) f ( (\varepsilon - E_F) k_B T )$.

The particle number $n_\odot$ is a function of the tunneling rates and still contains the temperature. 
However, for temperatures $k_B T \ll \alpha ( E_F )$ these become essentially independent of the temperature
such that we can concentrate on the temperature dependence of the rest. 
The integral with respect to $x$ contains the derivative of the Fermi function and is thus concentrated in a 
region of the order of one around zero. 
Therefore, we can in the case of small temperatures compared to $\alpha ( E_F)$, $k_B T \ll \alpha ( E_F),$ 
neglect the  linear term in $\phi_{a,b} (x)$ and estimate
\begin{displaymath}
 \phi_{a,b} (x) \approx g (x ) - a + log (b). 
\end{displaymath}
The simplification enables us to write
\begin{eqnarray*}
 G^{DSO} &\approx & 4 \kappa_l \kappa_{\bar l} \frac{e^2}{h}    \left( 2 - n_\odot \right)  \\   
 & & F^{DSO} \left( - \frac{E_{10} -             
{\bar E_{10}} }{\alpha(E_F)} +  \log \frac{T}{T_{\alpha(E_F)}} \right),
\end{eqnarray*}
with 
\begin{displaymath}
 F^{DSO} (c) = \int dx \frac { -\pi^2 f'(x) (1+f(x))  }  { \pi^2 ( 1 + f(x))^2 + 
( g(x) + c )^2   }.
\end{displaymath}
For large positive values of $c$, $F^{DSO} (c)$ takes small positive values. At some value ''$c_{max}``$ a maximum
is reached. In between there is a value of $c$, ''$c_{1/2}$``, where we have $F^{DSO}(c_{1/2})= 0.5 F^{DSO}
(c_{max})$.

We define the Kondo temperature by the demand $c = c_{1/2}$, i.e., 
\begin{eqnarray*}
 T_K &:=& e^{ c_{1/2} } \exp \left( \frac{E_{10} - E_F}{\alpha ( E_F) } \right)  \\
     & & T_{\alpha(E_F)} \exp \left( - p_{b^+_{T_{\alpha (E_F)}}}  (E_F) \right).   
\end{eqnarray*}
The second line seems to depend on the coupling $\alpha (E_F)$, but this dependence is weak because of the 
logarithmic dependence of  
$p_{b^+_T} ( E_F )$  on the temperature which we show in the appendix. The dependence on the bandwidth $W$ 
as introduced in 
figure \ref{figure explaining our choice of b} is proportionality as long as $\alpha(E_F) \ll W$ such that we 
arrive at
\begin{equation} \label{Kondo temperature} 
k_B T_K  = 7 W  \exp \left( \frac{E_{10} - E_F}{\alpha ( E_F) } \right) 
\end{equation}
after numerical evaluation of the constants. The prefactor, in our case ''7'', changes if shape of the band
(figure \ref{figure explaining our choice of b}) is chosen in a different way, for example, to be Lorentzian. 
The rest of the derivation is independent of the choice of the coupling function.

 \begin{figure}[h]\centering 
\includegraphics[width = 0.5\textwidth]{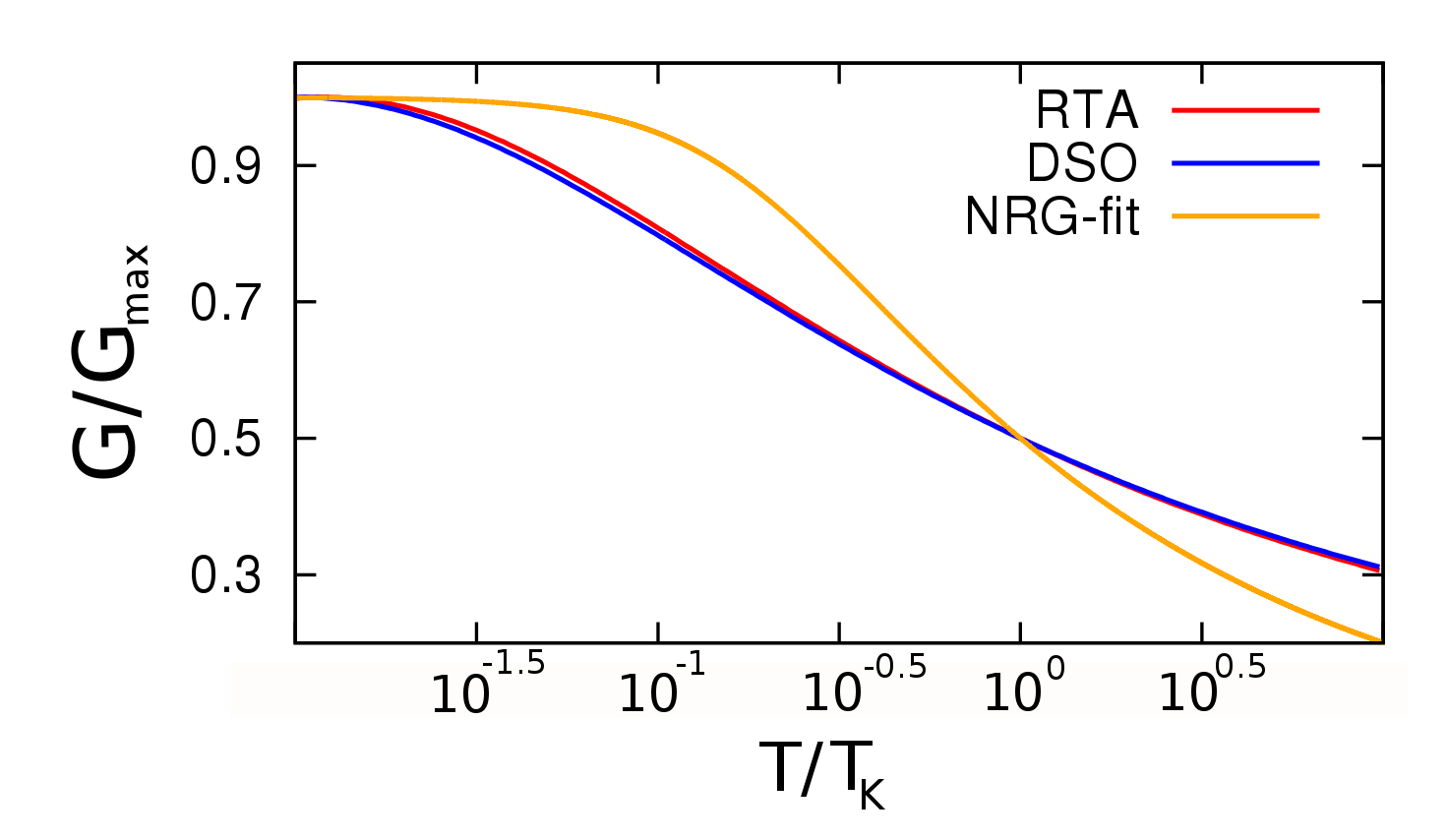}
\caption{\small  A comparison of the universal function of the DSO with an NRG-fit \cite{Goldhaber98}: Both 
functions take 
the value 0.5 at $T = T_K$ and are normalized in such a way that the maximum is one. A striking feature of the fit 
is that within one power of ten the linear conductance goes up from $50$ \% to about $95$ \% of its maximum while 
our function is growing much less in this interval. We included also the universal curve which one gets for 
the RTA. In agreement with figure \ref{G_RTA_and_DSO} and unlike the NRG-fit, the universal functions of RTA and 
DSO are going down again for even smaller exponents than shown here. The equation for the NRG-fit is 
$ G (T) = G_{max}  \left( 1/ \left[ (T/T_K')^2  + 1  \right]  \right)^s $ with $ T_K'= 
T_K /  ( 2^{1/s} - 1)^{1/2}  $ so that $G(T_K) =   G_{max} / 2$; we chose $s = 0.2$. }
\label{Comparison with NRG}
\end{figure}

We summarize the conditions for which the done simplifications are valid: 
\begin{displaymath}
 k_B T \ll \alpha (E_F ) \ll W .
\end{displaymath}
Then we can represent 
\begin{displaymath} 
 G^{DSO} = 4 \kappa_l \kappa_{{\bar l}} \frac{e^2}{h} ( 2 - n_\odot )  F^{DSO} \left( 
c_{1/2} + \log  \frac{T}{T_K}  \right),
\end{displaymath}
where $T_K$ is given by Eq. (\ref{Kondo temperature}).
The linear conductance becomes a universal function of $T/{T_K}$; at $T = T_K$, $G(T)$ reaches one half 
of its maximum. In figure \ref{Comparison with NRG} we compare our result for the universal function with 
a fit to the one obtained by NRG-calculations \cite{Costi94}. Essentially the same arguments one can apply to the 
RTA in order to obtain an analogous universality; the formula for $T_K$ deviates only in the prefactor. The 
relation between $\alpha (E_F)$ and the coupling parameter ``$\Gamma$`` by the use of which $T_K$ is most 
frequently expressed, e.g. \cite{Wingreen94, Schoeller97, Goldhaber98},  is   $ \Gamma = 2 \pi \alpha ( E_F ) $.

We acknowledge very clearly that the DSO fails to describe the regime of strong coupling quantitatively 
correctly. However, we think it is very remarkable that the linear conductance obtained by it displays a 
universality {\em in the same sense} as it is predicted by perfectly different theories.

\subsection{Zero bias anomaly of the differential conductance}
\begin{figure}[h]\centering 
\includegraphics[width = 0.5\textwidth]{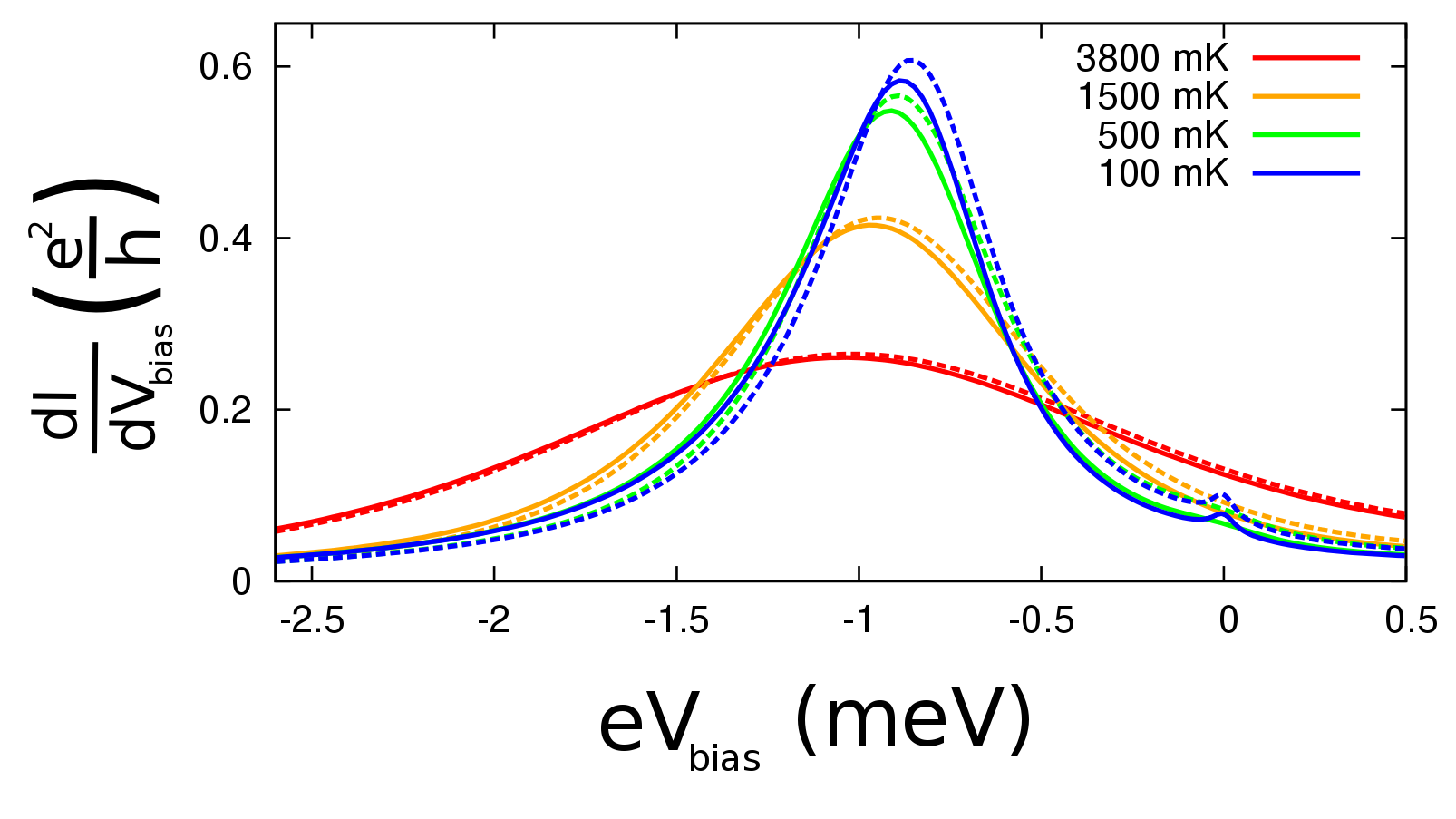}
\caption{\small (Colour online.) The differential conductance versus the bias. We set $E_{10}^{(0)} = E_F - 1meV$ 
and chose 
$ \alpha( E_F ) = 0.042 meV,  W = 1 eV  $ as in figure \ref{G_RTA_and_DSO}. We 
fixed one of the chemical potentials at the Fermi level, $\mu_{l_0} = E_F$, changed only the other one and 
defined $eV_{bias} = \mu_{\bar {l_0}} - E_F$. 
We assumed a capacitive 
coupling between the leads and the quantum dot in such a way that 
$E_{10}(V_{bias}) = E_{10}^{(0)}  + 0.2 eV_{bias} $.  We see a resonance appearing at zero bias for small 
temperatures. The dashed lines show the result of the RTA. The resonance is becoming more and more pronounced with 
decreasing temperature as it is observed in experiments \cite{Grobis08}.  The shape of the curve depends on the 
capacitive coupling and on how the window between the two chemical potentials is opened; however, the 
appearing of the zero bias anomaly does not in principle depend on these choices as one can conclude from the 
fact that they are irrelevant for the differential conductance at zero bias.}
\label{zero_bias_resonance}
\end{figure}

In addition to the linear conductance we considered the differential conductance obtained within the infinite-$U$ 
DSO.
We notice that in qualitatively the same way as the RTA \cite{König96} the approximation produces a zero bias 
maximum of 
the differential conductance in case $E_{10}$ lies below the Fermi level (Figure \ref{zero_bias_resonance}) and a 
minimum in case it lies above or in the vicinity of the Fermi level. The effect is getting more pronounced 
for smaller and smaller temperatures.

The generalization of the infinite-$U$ DSO to the case of different energies $E_{\uparrow} \neq E_{\downarrow}$
is straightforward. One obtains the tunneling rates
\begin{widetext}
 \begin{equation} \label{B_field_transition_rates}
\Gamma_{l \sigma}^\pm = \frac{2\pi}{\hbar} \int d \varepsilon \frac
{ \alpha_l^\pm (\varepsilon) \left( \alpha (\varepsilon ) + \alpha^+ (\varepsilon + E_{{\bar \sigma} \sigma})
      \right)  }
{\pi^2  \left( \alpha (\varepsilon) + \alpha^+ (\varepsilon + E_{{\bar \sigma} \sigma})  \right)^2  +
\left( \varepsilon - E_{\sigma 0} + p_{\alpha} (\varepsilon) + p_{\alpha^+} ( \varepsilon + 
E_{ { \bar \sigma } \sigma } )
\right)^2 }.
\end{equation}
\end{widetext}

\begin{figure}[h]\centering 
\includegraphics[width = 0.5\textwidth]{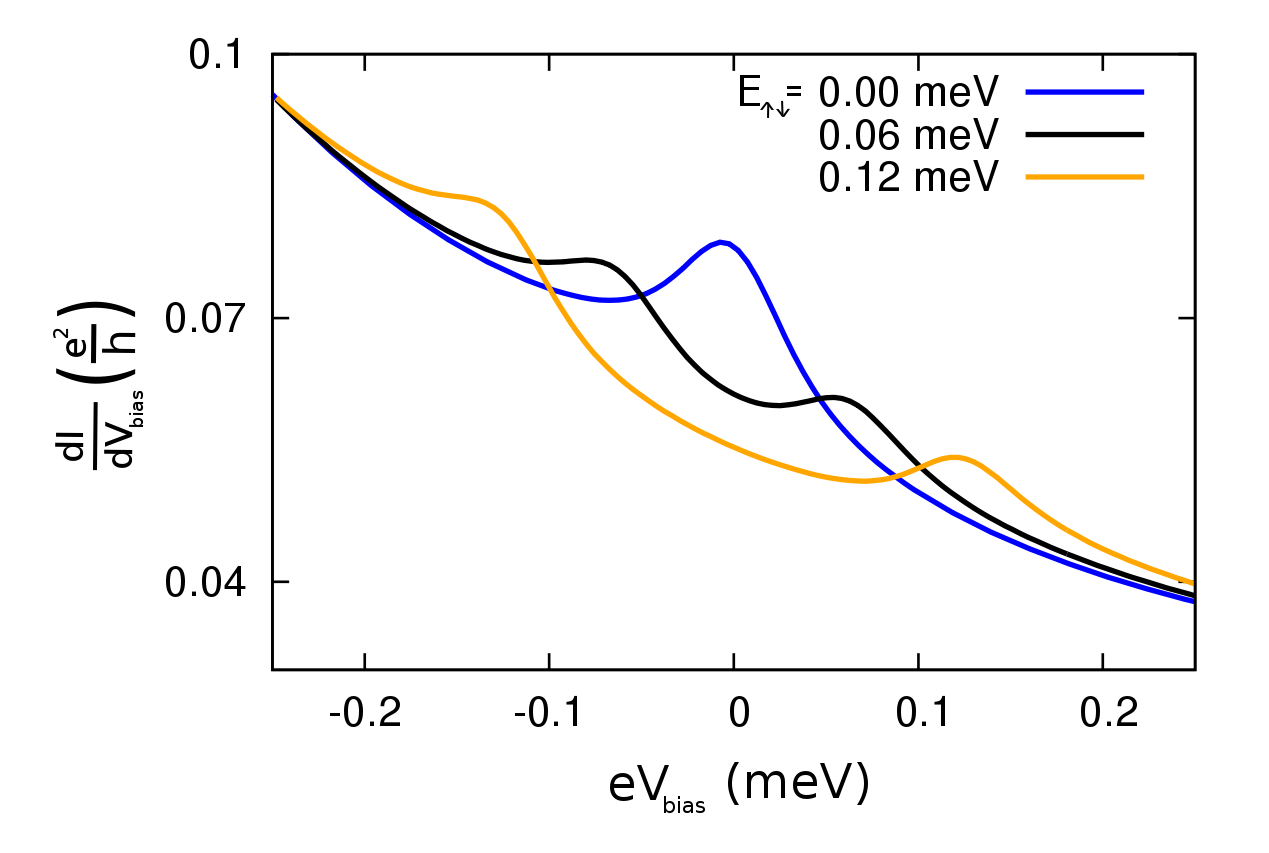}
\caption{\small  The zero bias resonance is split up if $E_\uparrow \neq E_\downarrow$. We chose the 
temperature $T=100mK$ and the remaining parameters as in figure \ref{zero_bias_resonance} apart from the 
splitting, 
$E_{\sigma 0}^{(0)} = E_F - 1meV \pm E_{ \sigma {\bar \sigma} }/2 $.  }
\label{Zeeman-splitting}
\end{figure}
We see (Figure \ref{Zeeman-splitting}) that the zero bias anomaly is split according to 
$ eV_{bias} \approx \pm E_{\uparrow \downarrow}$, in  agreement with theoretical results \cite{Meir93, König96}
and experiments \cite{Schmid98, Ralph94}.

\subsection{Situations in which only one resonance is expected}
\begin{figure}[h]\centering 
\includegraphics[width = 0.5\textwidth]{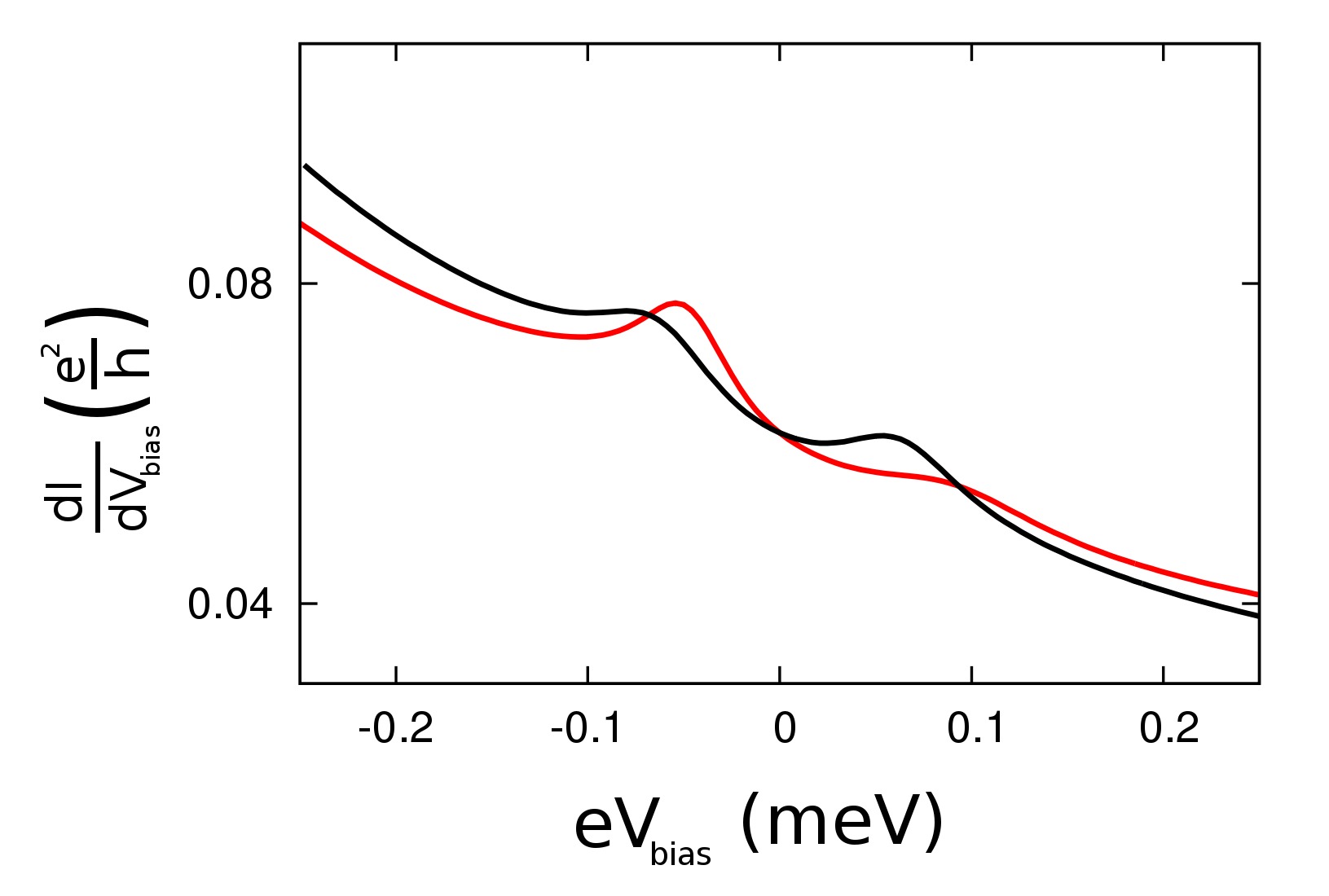}
\caption{\small  An asymmetry of the capacitive couplings of the levels to the leads has the effect that 
the resonances become asymmetric; one is starting to vanish, the other one is getting sharper. The black line 
is the one already appearing in figure \ref{Zeeman-splitting}. For the red (grey) line we changed the capacitive 
couplings in such a way that $E_{\uparrow 0}(V_{bias}) = E_{\uparrow 0}^{(0)}  + 0.4 eV_{bias} $, 
$E_{\downarrow 0}(V_{bias}) = E_{\downarrow 0}^{(0)}  + 0.1 eV_{bias} $. }
\label{capacitive_asymmetry}
\end{figure}
In Ref. \cite{Schmid98} a resonance close to zero bias whose position changed slightly with 
the gate voltage was reported. The dependence of the position on the gate voltage
was explained by the conjecture that two different wave-functions (not only two different spins) might be 
involved, such that  the assumption of different capacitive couplings of the levels to the gate 
electrode was justified. However, in this case one would expect to see a second peak at the opposite bias. 
This was not measured. We assume different capacitive couplings of the levels to the leads and obtain 
that with growing asymmetry one of the peaks is changing position, getting wider and much less pronounced. 
The other one, however, is getting sharper (figure \ref{capacitive_asymmetry}). The explanation for this 
behavior at 
the level of the transition rates, Eq. (\ref{B_field_transition_rates}), is that $\Gamma_{l\sigma}^\pm$ changes 
considerably with the bias in regions where 
\begin{displaymath}
 \mu_l - \mu_{{ \bar l }}  \approx  E_{\sigma {\bar \sigma} },      
\end{displaymath}
since then the region of large values of $p_{\alpha_l^+} (\varepsilon - E_{{\bar \sigma} \sigma}) $ is leaving
or entering the interval over which the integral essentially goes.  
This leads to the condition  ``$e V_{bias} \approx  \pm E_{\uparrow \downarrow}$'' for rapid change of the current 
with the bias.  
In case of different capacitive couplings of the levels to the leads the energy difference becomes a function
of the bias. With increasing bias, one of the differences is decreasing while the other one is increasing. 
Thus, one of the resonances is getting sharper while the other one is smeared out.  The positions
are no longer symmetric with respect to zero bias.

Moreover, we notice that also asymmetric tunnel coupling can have the effect that one of the resonances is 
getting less pronounced. One can let the coupling functions $\alpha_l (\varepsilon) $, Eq. 
(\ref{definition of alpha}),
be dependent on the spin as well as on the lead and thus obtain further independent parameters. 
We evaluated the differential conductance also in this case (not shown) 
and we can qualitatively confirm the assumption that different tunnel couplings of the levels to source and 
drain, too, can be responsible for the observation of only one peak \cite{Schmid98}.

\begin{figure}[h]\centering 
\includegraphics[width = 0.5\textwidth]{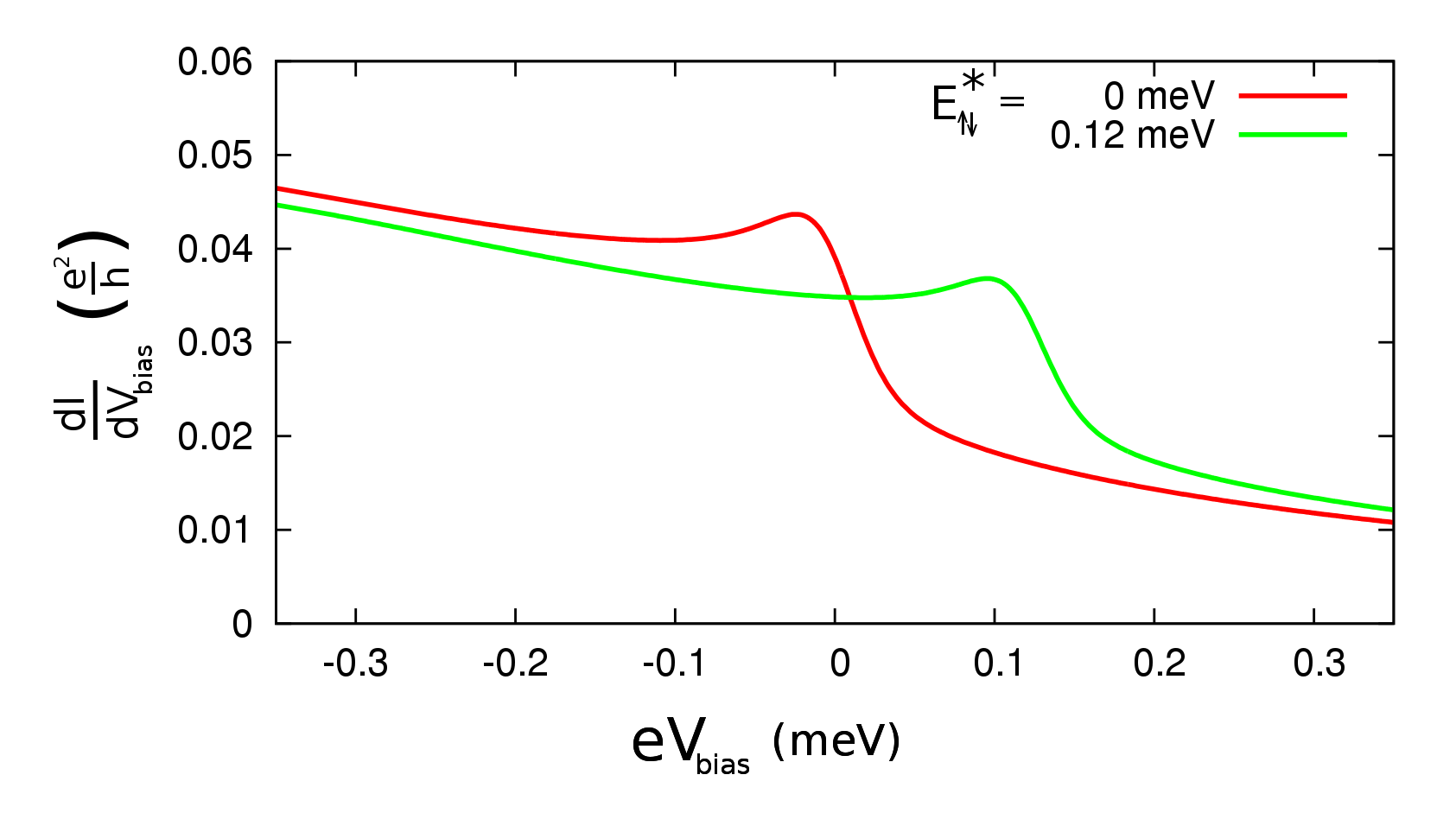}
\caption{\small The differential conductance as function of the bias in the following situation: The temperature 
is $T = 100 mK$, the capacitive coupling we chose to be zero; moreover, we 
chose $\mu_\downarrow =  E_F - 0.5 meV, \mu_{l_0 \uparrow} = E_F + 0.5 meV $. The energies are chosen as: 
$E_{\sigma 0} = \mu_{l_0 \sigma} - 1 meV + E_{\sigma {\bar \sigma}}^* / 2 $. }
\label{only one}                 
\end{figure}

As suggested in Ref. \cite{Report12}, we consider a  
second situation where the DSO yields, this time,  {\em in principle} only one resonance: The 
energies $E_\uparrow, E_\downarrow$ are different and there are four different, separately variable, chemical 
potentials $\mu_{l\sigma}$ for each of the leads and each of the spins. The chemical potentials of the 
down-spin are kept constant and equal, $\mu_{l \downarrow} =: \mu_{\downarrow} $; one of the up-spin 
chemical potentials, too,  is kept constant. 
Only $\mu_{\bar {l_0} \uparrow}$ is varied. The current is 
considered as a function of $eV_{bias} = \mu_{\bar {l_0} \uparrow} - \mu_{l_0 \uparrow}$. The DSO can be easily 
applied to such a situation. The initial density matrix of the contacts factorizes into four instead of two 
components. The coupling functions become spin-dependent. The tunneling rates of the infinite-$U$ DSO read in the 
most general case: 
\begin{widetext}
 \begin{equation} 
\Gamma_{l \sigma}^\pm = \frac{2\pi}{\hbar} \int d \varepsilon \frac
{ \alpha_{l \sigma}^\pm (\varepsilon) \left( \alpha_\sigma (\varepsilon ) + \alpha_{{\bar \sigma}}^+ 
(\varepsilon + E_{{\bar \sigma} \sigma})
      \right)  }
{\pi^2  \left( \alpha_\sigma (\varepsilon) + \alpha_{\bar \sigma}^+ (\varepsilon + E_{{\bar \sigma} \sigma})  
\right)^2  + \left( \varepsilon - E_{\sigma 0} + p_{\alpha_\sigma} (\varepsilon) + p_{\alpha_{{\bar \sigma}}^+} 
( \varepsilon + E_{ { \bar \sigma } \sigma } )
\right)^2 }.
\end{equation}
\end{widetext}
The rates $\Gamma_{l \sigma}^\pm $ change rapidly with the bias in regions where 
\begin{displaymath}
 \mu_{l\sigma} - \mu_{ l'{\bar \sigma} }  \approx  E_{\sigma {\bar \sigma}}.
\end{displaymath}
This leads to the resonance condition 
\begin{equation} \label{resonance condition}
eV_{bias} \approx  E_{\uparrow \downarrow}^*  := E_\uparrow^* - E_\downarrow^*, 
\end{equation}
where we used the definition $E_\sigma^* := E_{\sigma 0} - \mu_{l_0 \sigma}$.
Indeed, a plot of the differential 
conductance as function of the bias displays one resonance located approximately at this value of the 
bias (Fig. \ref{only one}). 
Experiments with a pseudo-spin\cite{Wilhelm02} might be interpreted by the use of the SIAM. The DSO predicts 
the appearing of only one resonance in case only one of the voltages is varied.

A further in principle possible experiment with pseudo-spin would be the following: The two differences 
$\mu_{l\uparrow} - \mu_{l\downarrow}, l = l_0, \bar{l_0},$ are held constant and equal. The two 
voltages $\mu_{\bar{l_0} \sigma} - \mu_{l_0 \sigma} =: eV_{bias}$ are equal and are varied. This corresponds 
so far to an experiment with real spin. The DSO yields two resonances at voltages 
$ eV_{bias} \approx \pm E_{\uparrow \downarrow}^*$. However, the current can be viewed as the sum of the two 
spin-currents, i.e., the current of the $\uparrow$-electrons plus the current of the $\downarrow$-electrons. 
An advantage of an experiment with pseudo-spin is that, in principle, the two currents can be measured 
separately. In the theory, anyway, it is not a problem to consider the two components separately. 
The application of the DSO yields in this situation two resonances of each of the two spin-components of the 
current. To conclude the discussion of the behavior of the resonance obtained by the DSO under various 
conditions, we can say that the predictions of the DSO are in agreement with those obtained by a purely 
qualitative approach\cite{Report12}. Apart from this, the predictions of the DSO are, up to our knowledge, 
novel.

\section{Linear conductance at finite $U$}
\begin{figure}[h]\centering 
\includegraphics[width = 0.5\textwidth]{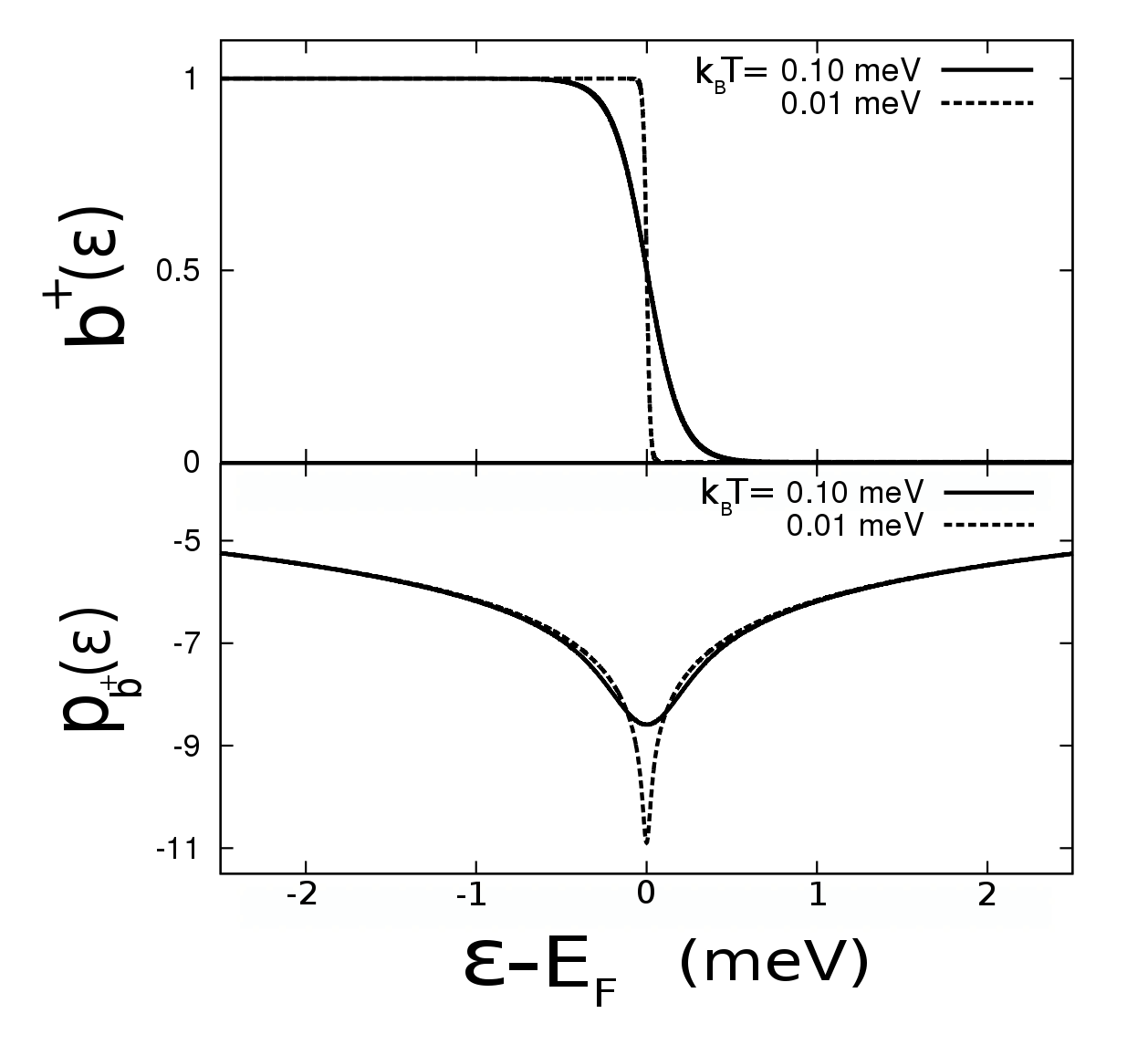}
\caption{\small  Behaviour of the normalized function $b_T^+:=\alpha^+(\varepsilon)/\alpha
(E_F)$ for which we use the abbreviation $b_T^+$ and the resulting behaviour of the function $p_{b_T^+}
(\varepsilon) := \int d\omega (b_T^+(\varepsilon + \omega) - b_T^+(\varepsilon - \omega))/\omega$ around the 
Fermi level.  
It measures the surplus of electrons with energy larger than $\varepsilon$ with a strong emphasis on the 
situation locally around $\varepsilon$. Therefore, the principal parts display a dip around $E_F$. 
If we decrease the temperature by a factor of ten, then the value of the principal part in the centre goes down 
by the logarithm of ten. Moreover, if we stretch the narrower of the two dips by this factor, then we obtain
the other dip. There is a well defined universal shape of the dips as we will show in the appendix.}
\label{behaviour of principal parts}
\end{figure}
In this section we investigate the linear conductance at finite interaction according to Eq. 
(\ref{linear conductance}).
By considering the integrals we can see qualitatively that we can expect an enhancement of the 
conductance with decreasing temperature if $E_{10}$ lies below and  $E_{21}$ lies above the Fermi level: 
The function $ \frac{-1}{k_B T} f' \left( \frac{ \varepsilon - E_F  } { k_B T  }    \right) $ of $\varepsilon$
has total weight one and is concentrated in a region of the size of the thermal energy around the Fermi level. 
In the denominators (Eq. (\ref{DSO transition rates})) the behaviour of the 
$p$-functions becomes important. 
The principal part $p_h (\varepsilon) $, Eq. (\ref{principal part}), measures an asymmetry of the function $h$ with 
respect to $\varepsilon$ ($h$ is an arbitrary function here). In particular, $p_{\alpha^+}$ takes negative values 
around the Fermi level since
in this region $\alpha^+ (\varepsilon)$ is decreasing (figure \ref{behaviour of principal parts}).

If we decrease the 
temperature, then the decay of 
the values of $\alpha^+$ will be more rapid and thus the absolute values of $p_{\alpha^+}(\varepsilon)$ are 
getting larger. 
Around the Fermi level, $\varepsilon + p_{\alpha^+} (\varepsilon)$ approaches the energy difference $E_{10}$
and the integral increases. At some point the sum even reaches and crosses the level position and then 
the integral decreases again. For the other integral the arguments are analogous. The energy correction
$p_{\alpha^-}$ is positive here and increases if we decrease the temperature.

\subsection{DSO-Conductance from weak to strong coupling}
\begin{figure}[h]\centering 
\includegraphics[width = 0.5\textwidth]{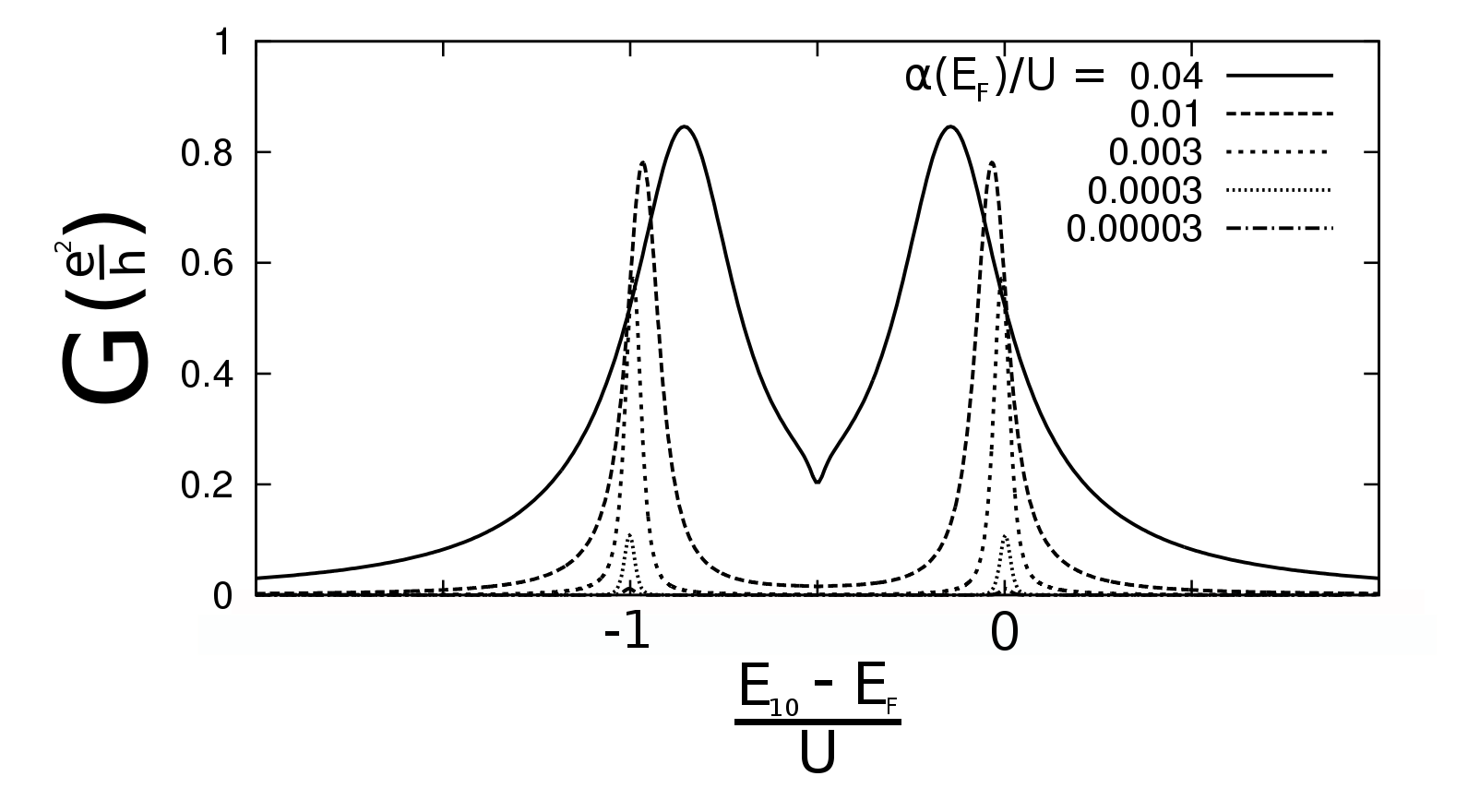}
\caption{\small  Plots of the differential conductance at zero bias as a function of $E_{10}$ (corresponding to 
a plot as a function of the gate voltage) for different tunnel couplings. We chose the interaction to be 
$U=1meV$ and the temperature $T=100 mK$. For weak coupling we see peaks of small height whose width is given by
the temperature; the peak positions are quite precisely defined by the resonance conditions $E_{10} = E_F$
and $E_{21} = E_F$. For increased coupling the corrections are becoming more and more important; the width is 
increasing with $\alpha (E_F)$, the peak position is shifted by the corrections
$p_{\alpha^\pm }$. The DSO approximation breaks for strong couplings where it produces a strange sharp dip in 
the centre of this plot which is not observed in reality.}   
\label{weak and strong coupling}
\end{figure}

For the numerical implementation of Eq. (\ref{linear conductance}) we wrote the coupling functions 
$\alpha (\varepsilon)$ still as 
$\alpha (E_F) b (\varepsilon)$, where the choice of $b(\varepsilon)$ is given in figure 
\ref{figure explaining our choice of b}. We  modeled the cases of strong and weak tunnel coupling
by large and small factors $\alpha (E_F)$, respectively. In figure \ref{weak and strong coupling} we show plots 
of $G^{DSO} (E_{10})$ for different values of the tunnel coupling. In the limit of weak coupling we reproduce 
the result of the second order theory while for increased coupling we expect essentially three effects: The peaks 
are getting  higher, broader and the maxima are moving towards each other. 

For strong coupling and for values of $E_{10} \approx E_F - U/2$ we expect irreducible 
tunneling processes outside the DSO-approximation to become more relevant: The population of all of the four 
possible states of the quantum dot is of the order of one here. Hence,  processes which 
transfer $|0\rangle\langle 0|$ into  $|2\rangle\langle 2|$ or vice versa or $|\sigma \rangle\langle \sigma|$ into  
$|{\bar \sigma}\rangle\langle {\bar \sigma}|$ should become more relevant. This might be an explanation for 
the failure of our diagram selection in describing this regime of parameters.

\subsection{From the empty orbital regime to the Kondo regime}
We want to compare now the result of the DSO for the linear conductance as a function of $E_{10}$ and of the 
temperature with experimental data \cite{Goldhaber98}. In the experiment a region within a two dimensional 
electron gas was isolated by electrostatically generated tunneling barriers. In this way a quantum dot 
which is tunnel coupled to leads was formed. Via a gate voltage it is possible to vary $E_{10}$.  
The linear conductance was measured as a function of the gate 
voltage and the temperature. The results were interpreted in terms of the SIAM. 
The authors distinguish between three different regimes of parameters, 
depending on whether the level position, i.e., $E_{10}$ is far below the Fermi energy (here the 
particle number is one, ``Kondo regime``), in the vicinity of the Fermi level (''mixed valence regime``) or
above the Fermi level (here the particle number is zero, ''empty orbital regime``). We tested the 
performance of the DSO  approximation by adjusting the parameters and comparing with the experimental data.

\label{later comparison}

\begin{figure}[h]\centering 
\includegraphics[width = 0.5\textwidth]{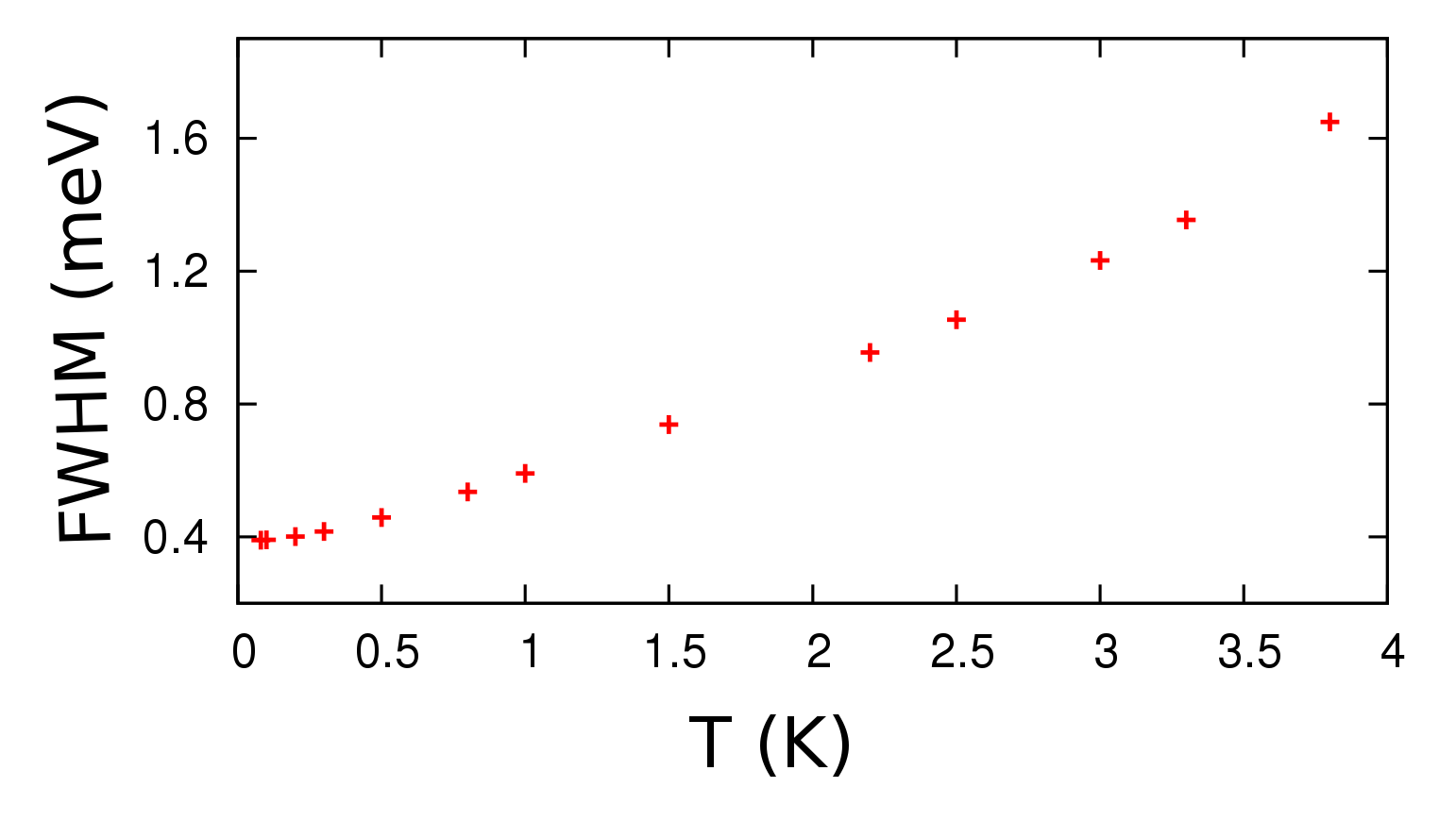}
\caption{\small A plot of the full widths at half maximum (''FWHM'') of the peaks in figure 
\ref{different temperatures} as a function of the temperature; the value of  $\alpha(E_F)$ is here $0.042meV$ . 
For small temperatures the FWHM seems to saturate  
at a value of about $0.39meV$ which is in agreement with the value of the experiment \cite{Goldhaber98}. 
For large temperatures the FWHM increases; the graph has positive curvature which 
is happening since the two peaks are getting mixed as we increase the temperature. This is why we cannot 
define the coupling constant in the same way as done in the experiment: There a linear dependence of the 
FWHM on the temperature for large temperatures was observed and the coupling constant was determined via 
the slope of the plot.}
\label{full widths at half max}
\end{figure}
We fitted the parameters in the following way: The temperatures are given explicitly. The value of 
the Coulomb interaction $U$, too, we take directly from the experiment. To fix the coupling parameter 
$\alpha(E_F)$ we plotted $G(E_{10})$ for various values  of it. 
We determined the parameter by the demand that the 
full width at half maximum of the peaks is close to the measured values 
(figure \ref{full widths at half max}).

\begin{figure}[h]\centering 
\includegraphics[width = 0.5\textwidth]{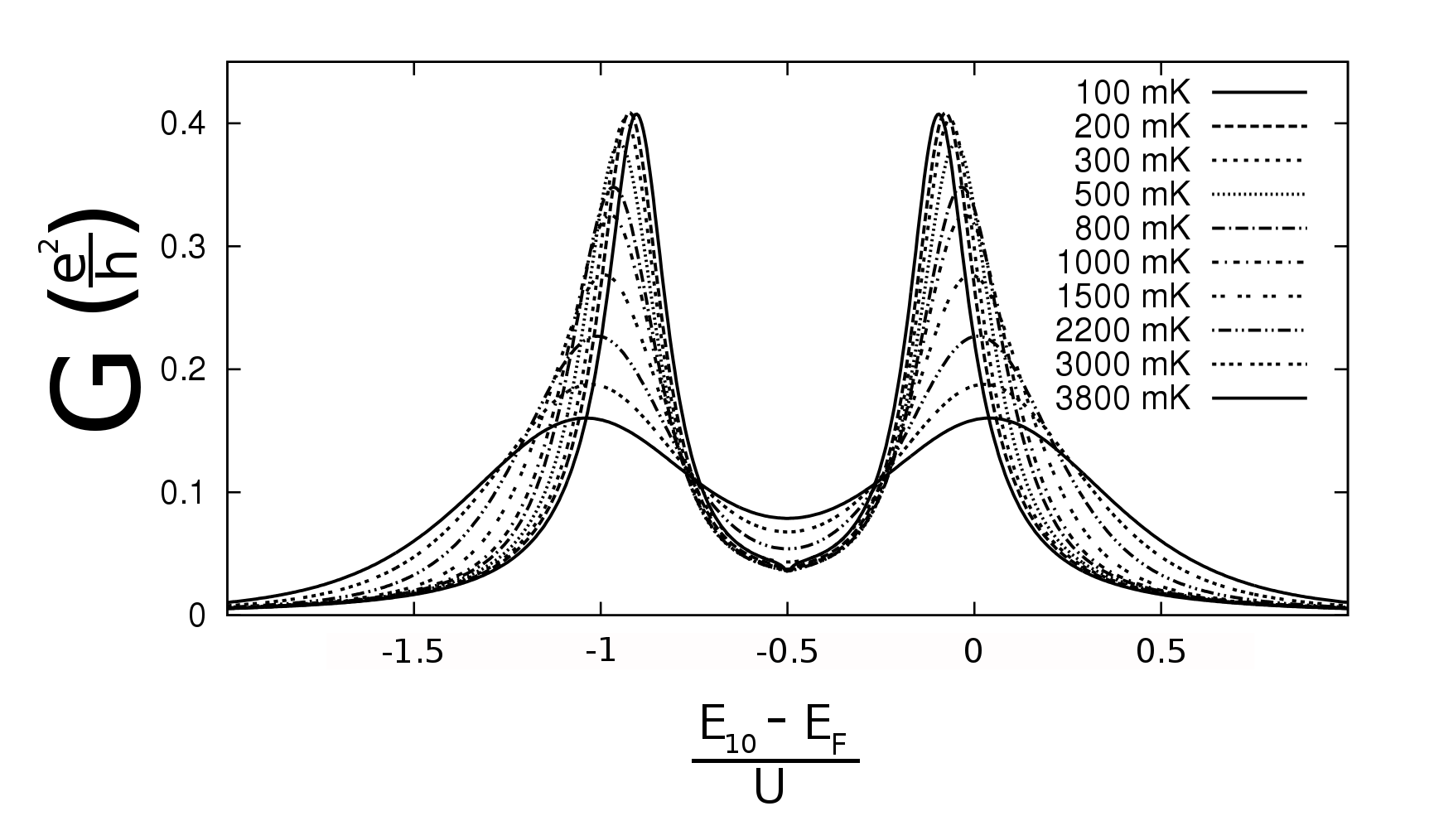}
\caption{\small A plot of the linear conductance as a function of the energy difference $E_{10}$ which corresponds
to a plot as a function of the gate voltage. We chose the interaction $U=1.9 meV$, the coupling $\alpha (E_F)
= 0.042 meV$, and the temperatures in 
agreement with the experiment \cite{Goldhaber98}; the asymmetry of the tunnel couplings to left and 
right lead we chose as $4\kappa_l \kappa_{\bar l} = 0.5$, i.e., we assume an asymmetry of about  
 $\kappa_l : \kappa_{\bar l} = 0.17$. For even smaller temperatures we would again see the sharp dip in the 
centre of the plots already shown in figure \ref{weak and strong coupling}. } 
\label{different temperatures}
\end{figure}
In figure \ref{different temperatures} we show a plot of the linear conductance as a function of the energy 
difference $E_{10}$ for different temperatures. We get qualitatively very similar 
behaviour as in \cite{Goldhaber98}. With decreasing temperature, 
the peaks are moving towards each other, they are getting higher and their widths are getting smaller and seem to 
saturate finally. In the end, we adjusted also the factor $4\kappa_l\kappa_{\bar l}$ which expresses  an  
asymmetry of the  tunneling couplings to source and drain. This, however, is only a fit made in such a way that 
the absolute values of the linear conductance are about the same in theory and experiment. The experiment 
was addressed also by Ref.\cite{Schoeller00} where a different asymmetry was assumed and good quantitative 
agreement was obtained. Therefore, it is 
difficult to formulate exact rules defining in what regime of parameters the DSO is quantitatively 
correct.

For a further comparison we show
the dependence of $G$ on the temperature for fixed $E_{10}$. In order to do this we 
express $E_{10}$ in terms of its position relative to the Fermi level and divide it by some quantity $\Gamma$
which characterizes the tunnel coupling in a similar way as in the experiment.    
\begin{figure}[h]\centering 
\includegraphics[width = 0.5\textwidth]{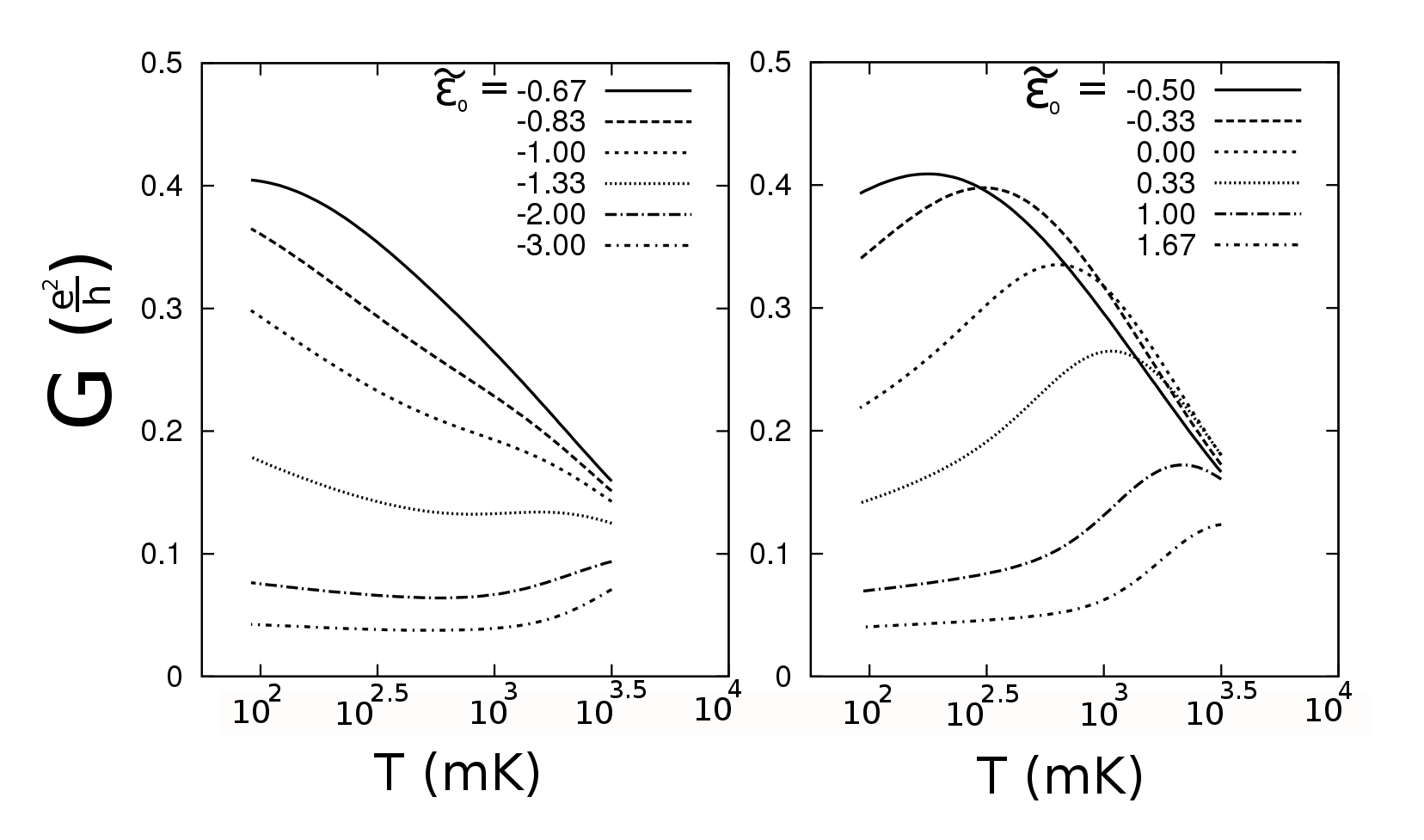}
\caption{\small A plot of $G(T)$ for fixed values of the gate voltage, expressed in terms of  
      ${\tilde {\varepsilon_0}} := (E_{10} -E_F)/\Gamma$. Here we take $\Gamma \approx 0.3 meV$ as in the   
    experiment with the argument that the saturation width of figure \ref{full widths at half max} is the same 
  as in the experiment. The different parameter regimes are called ''empty orbital regime`` 
  ($\tilde{ \varepsilon_0} > 0$), ''mixed valence regime`` ($-0.5 < \tilde{\varepsilon_0} < 0$) 
  and ''Kondo regime`` ($\tilde {\varepsilon_0} \ll -0.5$).  } 
\label{different gate voltages}
\end{figure}
By the plot we conclude that we have agreement of our theoretical result with 
the experimental data in the sense that the transition from the empty orbital regime, where with decreasing
temperature we see only a decrease and then rather constant behaviour, to the Kondo-regime, where we see only an 
increase, happens within an interval of $E_{10}$-values of the size of about $2\Gamma$.

On the other hand, we see a qualitative deviation in the regime $E_{10} \approx E_F$: The $G$ obtained by the 
DSO displays 
quite clearly a decrease with decreasing temperatures for small temperatures while in the experiment this 
decrease is much weaker. A study of the behaviour of the linear conductance in all three different regimes 
obtained by numerical renormalization-group calculations can be found in Ref. \cite{Costi10}.

\section{Conclusions}
We used a diagrammatic approach in order to describe transport across a SIAM quantum dot. We found 
a minimum selection of diagrams which we called ''dressed second order`` (DSO) diagrams, which 
straightforwardly yield the current in terms of transition rates. In general, the DSO represents the natural 
extension of the sequential tunneling approximation, valid for large interactions and when $k_B T \gg \Gamma$, 
to the regime $k_B T \sim \Gamma$. In particular, the conductance versus gate voltage exhibits peaks with a 
broadening no more given by the temperature but by the tunnel coupling.  
Appealing of the DSO is its simplicity and its potential for scalability to multilevel quantum dot systems. 
Moreover, its extension to set-ups with ferromagnetic or superconducting leads is straightforward. 

Furthermore, the diagram selection contains a zero bias anomaly developing at low temperatures. We showed that, 
if the degenerate level lies below the Fermi energy, then  it is a zero bias maximum of the differential 
differential conductance which appears for low temperatures and is getting more pronounced if the temperature
is decreased further. This is in qualitatively good agreement with experiments \cite{Grobis08, Schmid98}.

 We showed that the anomaly
displays features of the Kondo effect such as a universality in the dependence of the linear conductance as 
function of the temperature. We investigated the behaviour of the anomaly in case a magnetic field is applied
and discussed the impact of asymmetries with respect to capacitive or tunnel couplings to the leads. 
Moreover, we considered a situation where we expect in principle only one peak.  

We pointed out the close relation of the DSO diagram selection to that of the resonant tunneling
approximation (RTA) and compared their results in the case of infinitely large interaction. The RTA is more precise in 
the sense that it includes more diagrams; however, the DSO can be applied more easily to the case of 
energetically split levels. We showed that the inclusion of the diagrams outside the DSO which are contained 
in the RTA is not essential to the appearing of the zero bias anomaly. We think we found the 
smallest possible selection which contains the anomaly.  

Finally, we applied the approximation in the case of finite interaction and compared its result at the level 
of the linear conductance with an experiment, Ref. \cite{Goldhaber98}. We found good qualitative agreement with 
the experimental data. It is difficult to formulate general rules defining for what regime of parameters the 
DSO approximation is quantitatively correct. Being an extension of the sequential tunneling approximation, we 
expect the DSO to well cover the regime of large interactions down to temperatures $k_B T \sim \Gamma$.
Obvious failure of the approximation in describing the experiment \cite{Goldhaber98} we observed in the 
region $E_{10} \approx 1/2 ( E_F + E_F - U)$ for small temperatures.

The DSO for finite $U$ does not 
produce the observed plateau of the linear conductance as a function of the gate voltage for small temperatures
forming between the two resonances \cite{Grobis08}.
Additionally, the DSO for finite $U$ does not correctly describe the noninteracting, spin degenerate
limit. There is, however, 
a natural extension of the DSO which is indeed doing so, as outlined in Sec. \ref{spinless quantum dot}. Thus, 
there is hope that the DSO can be improved in such a way that the case of small interaction is described 
better. Moreover, the same class of diagrams could provide a natural way to extend the RTA to finite 
interaction as described by figure \ref{resonant tunneling}.       

In conclusion, the DSO is a novel approximation for the intermediate coupling regime which can additionally
provide useful insight also at temperatures $k_B T \ll \Gamma$.

\section{Acknowledgements}
We thank the DFG for financial support within the framework of the GRK 1570 and the SFB 689.

\section*{Appendix}
\label{appendix B}
We want to derive a representation 
\begin{eqnarray*}
 &&\int  d\varepsilon  \left. \frac { \pi^2 \alpha^2  (\varepsilon)  }
	{ d (\varepsilon)  } \right|^{V_{bias} = 0}  
          \frac{-1}{k_B T} f' \left( \frac{ \varepsilon - E_F  }
      { k_B T  }    \right)  \\ &\approx&  F^{RTA} \left( \frac{E_{10} - 
{\bar E_{10}} }{\alpha(E_F)}, \frac{k_BT}{\alpha(E_F)} \right),
\end{eqnarray*}
where the denominator of the integral is given by   $d( \varepsilon) := \pi^2 (\alpha + \alpha^+)^2 
(\varepsilon) + ( \varepsilon + p_{\alpha + \alpha^+} (\varepsilon) - E_{10} )^2$, the function 
$F^{RTA}$ is universal and where ${\bar E_{10}} $ does not depend on the gate voltage or on the temperature. 
To this end we write the second order function $\alpha (\varepsilon)$ as $\alpha (E_F) b (\varepsilon)$
and divide numerator and denominator of the integral by $\alpha(E_F)$. Moreover, we write 
$\varepsilon = E_F + x k_B T$ and integrate with respect to $x$ instead of $\varepsilon$. We argue then that the 
integral is concentrated in a region of a few multiples of the thermal energy around $E_F$ and that it is, 
because of this, 
for sufficiently small temperatures allowed to estimate $b(\varepsilon) = 1$ and $p_b (\varepsilon) 
= p_b (E_F)$;  $p_b (E_F)$ is zero because we chose the function $b$ to be symmetric around $E_F$. 
After these modifications we obtain the integral: 
\begin{equation} \label{technical integral}
 \int dx \frac { -\pi^2 f'(x)  }  { \pi^2 ( 1 + f(x))^2 + \phi^2 (x)  },
\end{equation}
where we used the abbreviation 
\begin{displaymath}
\phi (x) := x \frac{ k_B T } { \alpha(E_F) }  +  
\frac {E_F - E_{10}}{\alpha(E_F) } + p_{b_T^+} (E_F + x k_B T)
\end{displaymath}
and where the function $b_T^+$ is given by
$b_T^+ (\varepsilon) = b (\varepsilon) f \left(  (\varepsilon - E_F)/k_B T \right)$.

Thus, what remains to be done is the analysis of $p_{b^+} ( E_F + x k_B T)$. First of all, we consider its values
in $x = 0$ for different temperatures. By taking the derivative with respect to the temperature one obtains: 
\begin{eqnarray*}
 &&\frac{d}{dT} \int_0^\infty \frac{ b (E_F + \omega) f \left( \frac{\omega}{k_B T} \right) - 
   b (E_F - \omega) f \left( \frac{-\omega}{k_B T} \right) }  {\omega   } \\
&=& \frac{1}{T} \int_0^\infty dx -f'(x) \left[ b(E_F + x k_BT) +  b(E_F - x k_BT) \right] 
\\ & \approx & \frac{1}{T}. 
\end{eqnarray*}
For the final estimate we assumed that the temperature is sufficiently small such that in a region of a few
$k_B T$ around the Fermi level we have $b(\varepsilon) \approx 1$.

Secondly, we need to consider the values of $p_{b^+} ( E_F + x k_B T)$ for {\em one} temperature and different
values of $x$. We consider 
\begin{eqnarray} \label{what is replaced}
 && p_{b_T^+} ( E_F + xk_B T) - p_{b_T^+} ( E_F) = 
\end{eqnarray}
\begin{eqnarray*}
  & \int_0^\infty \frac{dy}{y} & 
  ( b( E_F + k_BT (x + y)) f(x + y)  \\
  && -  b(E_F + k_B T y) f(y)  \\
  && -  b(E_F + k_B T(x- y)) f(x-y)  \\
    && +  b(E_F - k_B T y) f(-y) ).  \\
\end{eqnarray*}
For every single value of $y$, the limit $T\to 0$ can be taken. We can guess that the limit of the integral 
is given by the integral of the point-wise limit, 
\begin{eqnarray} \label{definition of g}
g(x):=  & \int_0^\infty \frac{dy}{y} & 
   ( f(x + y)  -  f(y)   \nonumber \\
  && - f(x-y)       +  f(-y)). 
\end{eqnarray}
Then we replace the function of $x$ given by Eq. (\ref{what is replaced}) by $g(x)$ with the argument that 
for small temperatures the deviations between the two can be expected to be small. 

For the proof we want to apply Lebesgue's convergence theorem, so we need an integrable upper bound which is 
independent of the temperature. Moreover, we write $\int_0^\infty = \int_0^1 + \int_1^\infty$ since the different
intervals make different treatment necessary.

The integrand has the form 
\begin{displaymath}
(A B) (x + y) - (AB)(y) - (AB) (x-y) + (AB)(-y). 
\end{displaymath}
As to the integral $\int_0^1$, we group the terms with equal ''$x$'' into pairs and consider the two resulting 
differences separately. By adding and subtracting the mixed terms $A(x+y)B(x-y)$ one can see that
we have even a constant upper bound within this interval. The conditions which we demand from the function 
$b(\varepsilon)$ for this are the following: 
\begin{itemize}
 \item It is bounded, $\left| b(\varepsilon) \right| \le B$, $B$ independent of $\varepsilon$.
 \item It satisfies a Lipschitz condition of the form:  $\left| b(\varepsilon) - b(\varepsilon') \right|  \le
L \left| \varepsilon  - \varepsilon' \right| $, $L$ independent of $\varepsilon$ and $\varepsilon'$. 
\end{itemize}
 We mention that the Fermi function, too, has the two properties; the latter can be seen by 
using the fact that the derivative of the Fermi function is bounded and the mean value theorem. 
Moreover, a Lorentzian or our choice of the function $b$ (figure \ref{figure explaining our choice of b}) 
fulfills these conditions.

As to the integral $\int_1^\infty$, we group the terms with equal sign in front of $y$ into pairs. Again, it is
useful to add and subtract the mixed terms, e. g., $A(x+y)B(y)$. We obtain then: 
\begin{eqnarray}
  \frac{1}{y} b(E_F + k_BT(x +y)) \left( f(x +y) - f(y) \right) + && \nonumber \\
  \frac{1}{y} \left(  b(E_F + k_BT(x +y)) -  b(E_F + k_BT y)   \right) f(y). && \label{going to zero}  
\end{eqnarray}
 The point-wise limit of the first line is $1/y \left( f(x+y) - f(y)  \right)$, and the convergence is bounded
by $B/y \left|  f(x +y) - f(y) \right|$. This is integrable because we can estimate: 
\begin{eqnarray*}
 \frac{  \left|  f(x +y) - f(y) \right| } { y } &\le& 
max\left\lbrace   \left| f'(z) \right| : \left| z - y  \right| \le |x| \right\rbrace \left| x \right| \\   
&&    =: m_x (y)  \left| x \right| ,
\end{eqnarray*}
where we used $y \ge 1$.
We treat $x$ as a constant during these considerations. Because of the rapid decay of the derivative of the 
Fermi function $m_x (y)$ is integrable.

The point-wise limit of the second line is zero. In order to get an integrable upper bound we introduce
the function
\begin{displaymath}
 L_> (\varepsilon) := sup \left\lbrace  \frac {\left| b(\varepsilon'') - b(\varepsilon') \right|  } 
{ \varepsilon'' - \varepsilon'   }  : \varepsilon \le \varepsilon' < \varepsilon''
\right\rbrace.
\end{displaymath}
Then we can estimate the second line by
\begin{eqnarray*}
 && \left| \frac{1}{y} \left(  b(E_F + k_BT(x +y)) -  b(E_F + k_BT y)   \right) f(y) \right| \\
 &\le&   \frac{\left| x \right| }{y^2}  \left[ E_F +  k_B T (y - |x|) - E_F + k_B T |x|    \right]
  \\ && L_> ( E_F +  k_B T (y - |x|)),   
\end{eqnarray*}
where we multiplied and divided everything by $k_B T \left| x \right| y$. (The square bracket is just a 
complicated way of writing
``$k_B T y$``.) We note now that for sure $L_> (\varepsilon) \le L$ and 
make the further assumption that the function $\varepsilon L_> (\varepsilon) $ is bounded over any interval
which has a lower bound, i.e., for any $\varepsilon_0$ we have
\begin{equation}
 L^> (\varepsilon_0) := sup \left\lbrace  \left| \varepsilon L_> (\varepsilon) \right|: 
\varepsilon \ge \varepsilon_0  \right\rbrace < \infty.  \label{further assumption about b}  
\end{equation}
This assumption is fulfilled both for Lorentzian shapes of $b(\varepsilon)$ and for our way of choosing the 
second order function (figure \ref{figure explaining our choice of b}), the reason being the rapid decay of the 
derivatives of these functions. Using these properties we obtain for temperatures smaller than some arbitrary
temperature $T_0$ the upper bound: 
\begin{displaymath}
 \frac{\left| x  \right| }{y^2} \left\lbrace  L \left(  |E_F| + |x| k_B T_0  \right) + L^> \left(
  E_F - |x| k_B T_0  \right)   \right\rbrace. 
\end{displaymath}
 This is integrable with respect to $y$ over the interval between one and infinity and independent of the 
temperature between zero and $T_0$. With Lebesgue we can conclude that the integral of the function 
(\ref{going to zero}) of $y$  
really goes to zero. 
The terms with a minus in front of $y$ we can treat in the same way. For this we introduce functions 
$L_< (\varepsilon)$ and $L^< (\varepsilon)$ in analogy to the above method and demand the corresponding 
property of $b(\varepsilon)$ of assumption (\ref{further assumption about b}).

We showed the convergence 
\begin{displaymath}
 p_{b^+_T} ( E_F + k_BT x) - p_{b^+_T} (E_F)  \to g(x)  \quad\ (T \to 0)
\end{displaymath}
\begin{figure}[h]\centering 
\includegraphics[width = 0.5\textwidth]{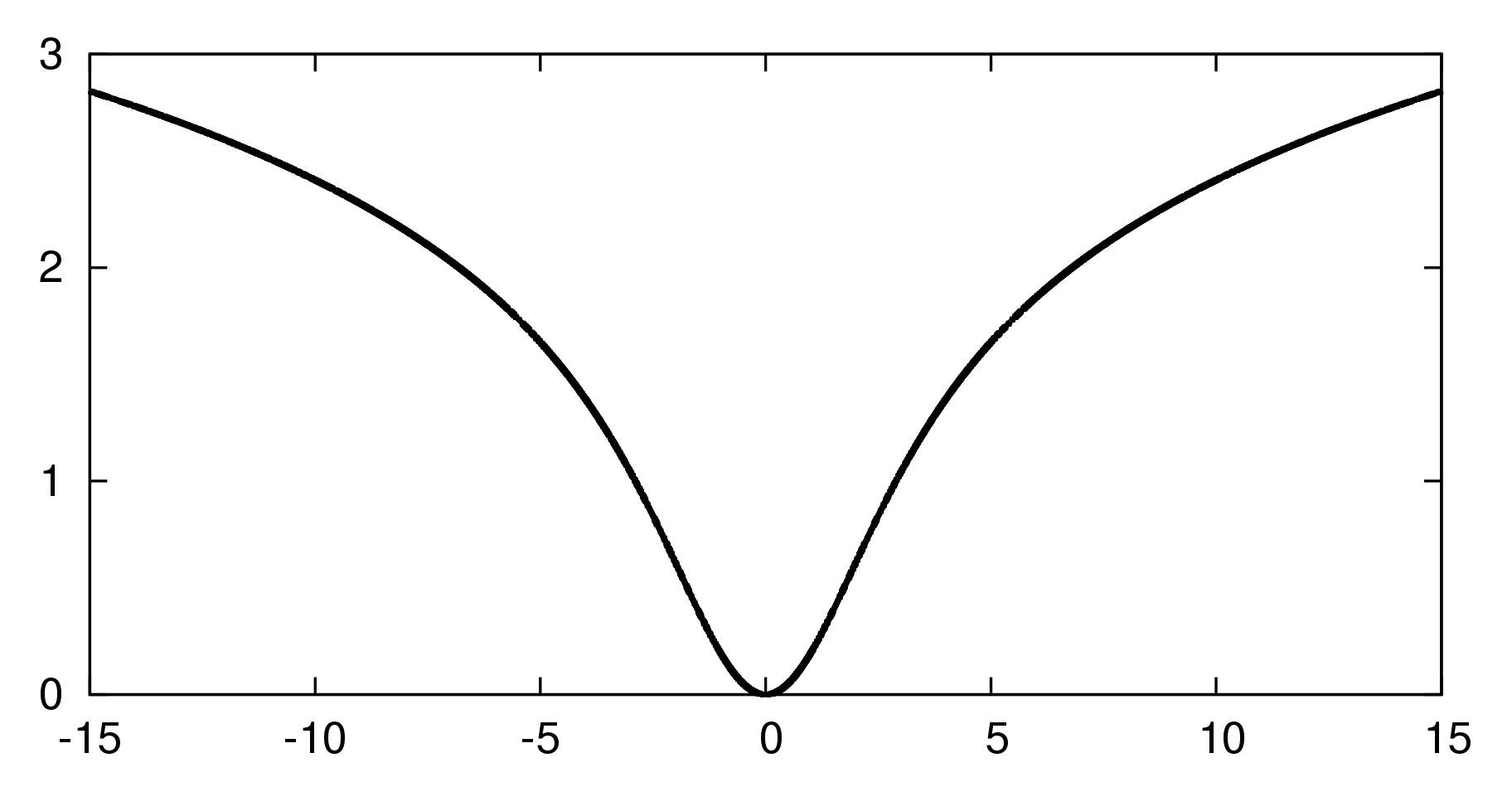}
\caption{\small A plot of the function $g(x)$ defined by eq. (\ref{definition of g}), i.e., the normalized shape 
of the functions $p_{b^+_T} (\varepsilon)$ around the Fermi 
level in units of the thermal energy. The growth of $g(x)$ is logarithmic in the sense that $xg'(x) \to 1  
\quad (|x| \to \infty)$. However, because of the presence of the derivative of the Fermi function in the integrals,
the behaviour of $g(x)$ around $x=0$ is more important for us. }
\label{plot of g}
\end{figure}
for arbitrary $x$, where the limit is given by the definition (\ref{definition of g}). 
 In figure \ref{plot of g} we  plotted  the function $g(x)$. Already earlier we noticed that the dependence of 
$p_{b^+_T} (E_F)$ on the temperature is logarithmic. By putting the two pieces of information together we can 
estimate: 
\begin{displaymath}
 p_{b^+_T} (E_F + x k_BT) \approx p_{b^+_{T_{\alpha(E_F)}}}          
(E_F) + log\left( \frac{T}{T_{\alpha(E_F)}} \right) + g(x),
\end{displaymath}
 where $T_{\alpha(E_F)}$ is defined by the condition $k_BT_{\alpha(E_F)} = \alpha(E_F)$. We insert this into the
integral (\ref{technical integral}) and obtain:   
\begin{eqnarray*}
\phi (x) &\approx& x \frac{T } { T_{\alpha(E_F)} }  + log\left( \frac{T}{T_{\alpha(E_F)}} \right)  + g(x)\\ 
&& + \frac {E_F - E_{10}}{\alpha(E_F) } + p_{b^+_{T_{\alpha(E_F)}}}   (E_F)  .
\end{eqnarray*}
Now we define a reference value for $E_{10}$, ``$ {\bar E_{10}}$'', by the
demand that the value of the second line is zero for $E_{10} =  {\bar E_{10}}$. The integral 
(\ref{technical integral}) has then the form 
\begin{displaymath}
 F^{RTA} \left( \frac{E_{10} - {\bar E_{10}}   } { \alpha{E_F}}, \frac{k_BT}{\alpha(E_F)} \right),
\end{displaymath}
 where the definition of $F^{RTA}$ is 
\begin{displaymath}
 F^{RTA} (a,b) = \int dx \frac { -\pi^2 f'(x)  }  { \pi^2 ( 1 + f(x))^2 + 
\phi_{a,b}^2 (x)  }
\end{displaymath}
with 
\begin{equation}  \label{definition of phi_ab}
 \phi_{a,b} (x) =   g(x) - a + xb +  log(b).   
\end{equation}

The corresponding integral in the formula for the linear conductance within the DSO in the infinite-U case  
can be represented in an analogous way. 
The difference is that in the numerator we get $-\pi^2 f'(x) (1 + f(x))$ instead of only $-\pi^2 f'(x)$.

\end{document}